\documentclass[
 reprint,
twocolumn,
superscriptaddress,
longbibliography,
 amsmath,amssymb,
 aps,
prb,
]{revtex4-2}

\usepackage{braket}
\usepackage{physics}
\usepackage{comment}
\usepackage{xcolor}
\usepackage{appendix}
\usepackage{caption}
\usepackage{subcaption}
\usepackage{graphicx}
\usepackage{dcolumn}
\usepackage[mathscr]{euscript}
\usepackage[scr=boondox]{mathalpha}
\usepackage{bm}
\usepackage[linktocpage]{hyperref}
\hypersetup{colorlinks=true,citecolor=blue,linkcolor=blue, urlcolor=blue, breaklinks=true}

\makeatletter
\newcommand\xlabel[2][]{\phantomsection\def\@currentlabelname{#1}\label{#2}}
\makeatother

\newcommand{\Z}{\mathbb{Z}}
\newcommand{\OO}{\text{o}}

\newcommand{\Cmop}{\tilde{C}_{M_{\text{o}}}^+}
\newcommand{\Cmom}{\tilde{C}_{M_{\text{o}}}^-}
\newcommand{\Cmopm}{\tilde{C}_{M_{\text{o}}}^{\pm}}
\newcommand{\MO}{M_\text{o}}

\begin{document}

\title{Complete crystalline topological invariants from partial rotations in (2+1)D \\ invertible fermionic states and Hofstadter's butterfly}

\author{Yuxuan Zhang, Naren Manjunath, Ryohei Kobayashi, Maissam Barkeshli}
\affiliation{Department of Physics, Joint Quantum Institute, and Condensed Matter Theory Center, University of Maryland,
College Park, Maryland 20742, USA}

\begin{abstract}
The theory of topological phases of matter predicts invariants protected only by crystalline symmetry, yet it has been unclear how to extract these from microscopic calculations in general. Here we show how to extract a set of many-body invariants $\{\Theta_\OO^{\pm}\}$, where $\OO$ is a high symmetry point, from partial rotations in (2+1)D invertible fermionic states.  Our results apply in the presence of magnetic field and Chern number $C \neq 0$, in contrast to previous work.
$\{\Theta_{\OO}^{\pm}\}$ together with $C$, chiral central charge $c_-$, and filling $\nu$ provide a complete many-body characterization of the topological state with symmetry group $G = \text{U}(1) \times_\phi [\mathbb{Z}^2 \rtimes \mathbb{Z}_M]$. Moreover, all these many-body invariants can be obtained from a single bulk ground state, without inserting additional defects.
We perform numerical computations on the square lattice Hofstadter model. Remarkably, these match calculations from conformal and topological field theory, where $G$-crossed modular $S, T$ matrices of symmetry defects play a crucial role. Our results provide additional colorings of Hofstadter's butterfly, extending recently discovered colorings by the discrete shift and quantized charge polarization.
\end{abstract}

\maketitle

There has been intense work over the past decade in developing a comprehensive characterization and classification of topological phases of matter with internal \cite{Kitaev2009periodic,ryu2010,Chen2013,kapustin2014SPTbeyond,kapustin2015fSPT,Wang2020fSPT,Senthil2015SPT,barkeshli2019,barkeshli2021invertible,aasen2021characterization,bulmashSymmFrac} and crystalline symmetries \cite{Chiu2016review,cheng2016lsm,Bradlyn2017tqc,Po2017symmind,Kruthoff2017TCI,Huang2017,Thorngren2018,cheng2018rotation,zhang2020realspace,Elcoro2021tqc,manjunath2021cgt,manjunath2020FQH}. Despite much progress, the problem of how to fully characterize such systems, in particular by numerically extracting a complete set of invariants, remains partially solved \cite{shiozaki2017invt,Shapourian2017FSPT,zhang2022fractional,zhang2022pol,herzogarbeitman2022interacting}.

In this paper, we study (2+1)D topological states of fermions with symmetry group $G = \text{U}(1) \times_\phi [\mathbb{Z}^2 \rtimes \mathbb{Z}_M]$, which consists of $\text{U}(1)$ charge conservation, discrete magnetic translation and point group rotation symmetries. We focus on invertible fermionic phases, which have a unique ground state on all manifolds \cite{barkeshli2021invertible,aasen2021characterization}. We show how to numerically extract a set of many-body invariants $\{\Theta^{\pm}_{\OO}\}$ associated to the crystalline symmetries by computing expectation values of partial rotations centered at high symmetry points $\OO$ of the unit cell. For example, if $\OO$ is preserved by rotations of order $M_{\OO}$ where $M_{\OO}$ is even, then for a fixed Chern number, $\Theta^{+}_{\OO}$ and $\Theta^{-}_{\OO}$ define $\Z_{M_{\OO}/2}$ and $\Z_{2M_{\OO}}$ invariants respectively, for each $\OO$. Our numerical results are in remarkable agreement with analytical calculations from conformal and topological field theory; specifically, the invariants $\{\Theta^{\pm}_{\OO}\}$ encode the $G$-crossed modular $S,T$ matrices of symmetry defects \cite{barkeshli2019,manjunath2020FQH}.

 \begin{figure}[t]
    \centering
    \includegraphics[width=7cm]{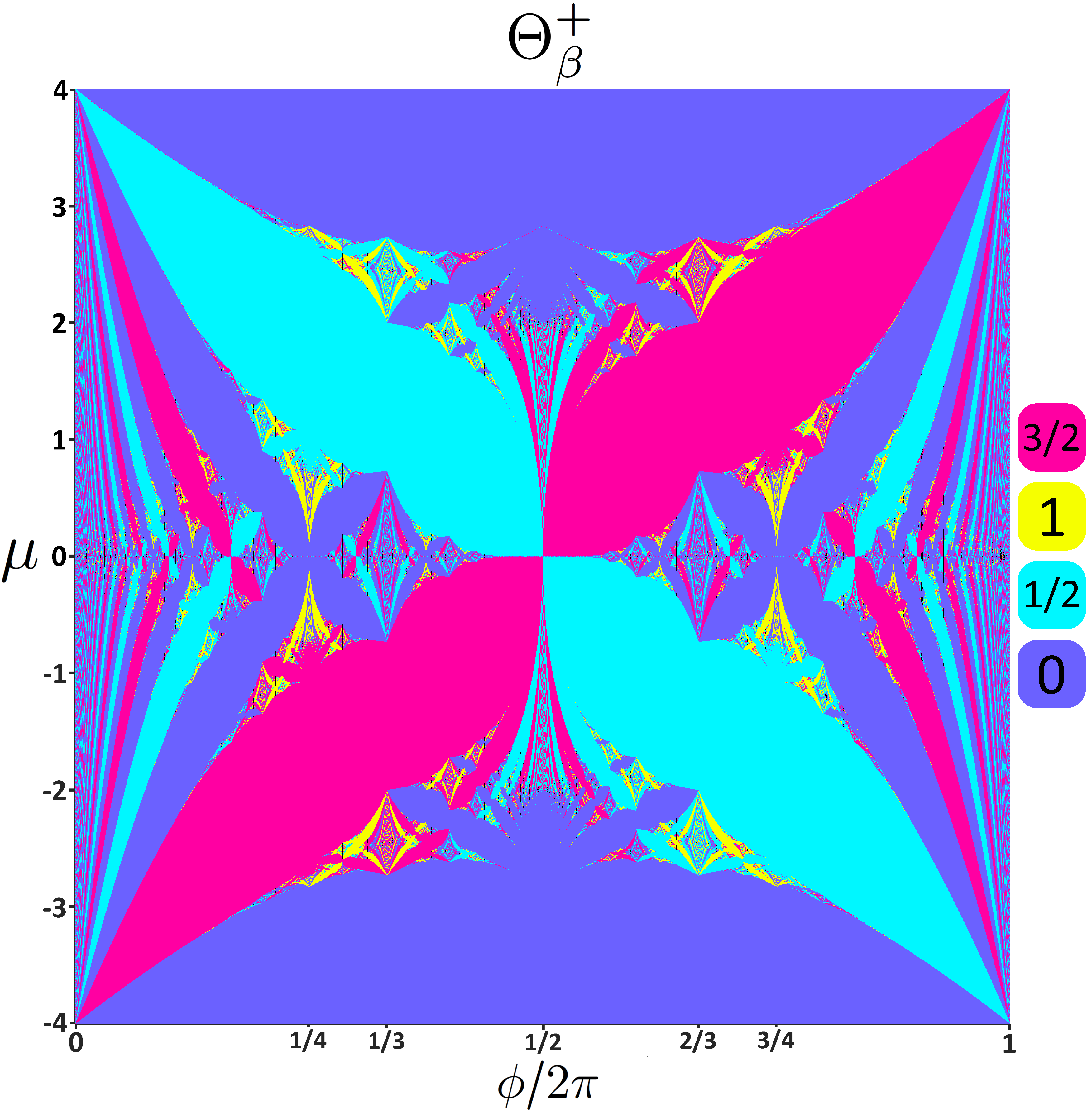}
    \caption{$\Theta_{\beta}^+ \mod 2$ for Hofstadter model, plotted using an empirical formula Eq.~\eqref{eq:Theta_full} and Eq.~\eqref{eq:theta_beta} in App.~\ref{app:num}.}
    \label{fig:beta_plus}
\end{figure}

 \begin{figure*}[t]
    \centering
    \includegraphics[width=16cm]{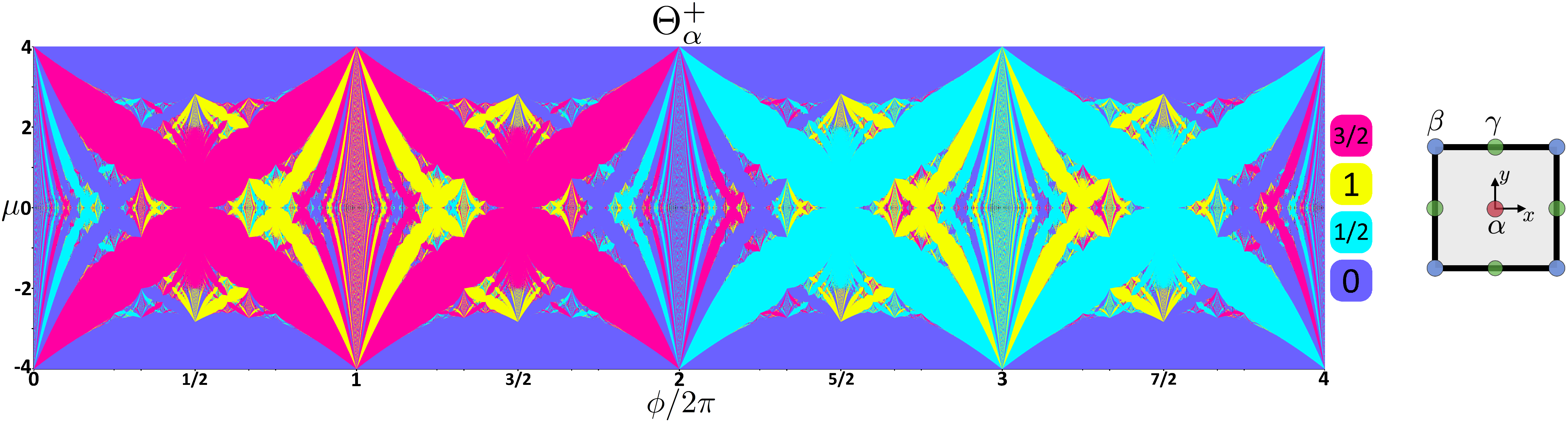}
    \caption{$\Theta_{\alpha}^+ \mod 2$ for Hofstadter model, where $\alpha$ is a plaquette center, plotted using an empirical formula Eq.~\eqref{eq:Theta_full} and Eq.~\eqref{eq:theta_alpha} in App.~\ref{app:num}. The choice of square lattice unit cell is shown on the right.}
    \label{fig:alpha_plus}
\end{figure*}

For square lattice and $\text{U}(1)$ charge conservation symmetries, the expected theoretical classification of topological invariants is given by the group $\Z^3\times\Z_8\times \Z_2\times \Z_4^2$ \cite{zhang2020realspace,manjunath2023classif}. 
We find that the invariants $\{\Theta^{\pm}_{\OO}\}$, together with the Chern number $C$, the chiral central charge $c_-$ and the filling $\nu$, give a \textit{complete} characterization of these invariants. In particular, the discrete shift $\mathscr{S}_{\OO}$ and quantized charge polarization $\vec{\mathscr{P}}_{\OO}$ studied in Refs.~\cite{zhang2022fractional,zhang2022pol} can be obtained from $\{\Theta^{\pm}_{\OO},C,c_-,\nu\}$. Our methods to extract $\{\Theta_{\OO}^{\pm}\}$ require only a single bulk ground state on a disk, without needing to insert additional defects, adding to a line of work that extracts topological invariants from single bulk ground states \cite{levin2006,kitaev2006topological,shiozaki2017invt,dehghani2021,cian2021,cian2022extracting,Kim2022ccc,fan2022}. 

As an application, we study the square lattice Hofstadter model and obtain a number of additional colorings of Hofstadter's butterfly, extending the colorings by $\mathscr{S}_{\OO}$ and $\vec{\mathscr{P}}_{\OO}$ discovered in \cite{zhang2022fractional,zhang2022pol}. Analogous classification results hold for point group rotations of order 2,3 and 6, and will be mentioned at the end of the paper.

Partial rotations and partial translations on a cylinder have been studied in prior works and have been shown to yield invariants such as topological spins of anyons, central charges~\cite{Qi2012momentumpolarization, FQHEDMRG, kobayashi2023extracting}, and invariants of symmetry-protected topological (SPT) states~\cite{Zaletel2014bosonicSPT,shiozaki2017invt, Shiozaki2018antiunitary, kobayashi2019,You2020hoe}. In contrast our results apply both in the presence of arbitrary rational magnetic flux, $C \neq 0$, demonstrate a complete set of invariants for crystalline symmetries, and also incorporate several subtleties missed in prior work. These include the proper use of $G$-crossed modular matrices,  the quantization of $\{\Theta^{\pm}_{\OO}\}$, and their dependence on $\OO$.

\paragraph*{Symmetries and model.} We study gapped phases of matter with the symmetry group $G = \text{U}(1) \times_{\phi} [\mathbb{Z}^2 \rtimes \Z_M]$, where $\Z^2$ denotes magnetic lattice translations, $\Z_M$ for $M = 2,3,4,6$ denotes point group rotations. We assume $M=4$ throughout, and briefly comment on $M=2,3,6$ at the end of the paper.

The symbol $\times_{\phi}$ implies that the magnetic translation operators, generated by $\tilde{T}_{\bf x}, \tilde{T}_{\bf y}$, obey the algebra $\tilde{T}_{\bf y}^{-1} \tilde{T}_{\bf x}^{-1} \tilde{T}_{\bf y} \tilde{T}_{\bf x} = e^{i \phi \hat{N}}$ where $\hat{N}$ is the total fermion number. The tilde indicates that the definition of the operator involves a $\text{U}(1)$ gauge transformation. Note $e^{i \pi \hat{N}} = (-1)^F$, which is fermion parity. 

Let $2\pi/M_{\OO}$ be the smallest angle of rotation which preserves the rotation center $\OO$. The possible high symmetry points $\OO$ of the square lattice are shown in Fig.~\ref{fig:alpha_plus}; they are $\alpha, \beta$ (plaquette center and vertex respectively, with $M_{\alpha} = M_{\beta}=4$) and $\gamma$ (edge center, with $M_{\gamma}=2$)~\footnote{Note we will abuse language somewhat and use $\OO$ both as the rotation center and as a maximal Wyckoff position.}. As shown in \cite{zhang2022fractional,zhang2022pol} and reviewed in App.~\ref{app:num}, there is a canonical choice of magnetic point group rotation operators $\Cmop$ which are centered at $\OO$ and satisfy $(\Cmop)^{\MO} = +1$. We also define another set of operators 
$\tilde{C}_{M_\OO}^- := e^{i \frac{\pi}{M_{\OO}} \hat{N}} \tilde{C}_{M_{\OO}}^+$,
and $\tilde{C}_{M_{\OO}, \chi}^{\pm} := e^{i\chi \frac{2\pi}{\MO} \hat{N}} \tilde{C}_{M_{\OO}}^\pm$. Note that
$\tilde{C}_{M_{\OO}}^- = \tilde{C}_{M_{\OO}, 1/2}^+$ and $(\Cmom)^{\MO} = (-1)^F$. 

We study the spinless free fermion Hofstadter model with flux $\phi$ per unit cell, with Hamiltonian
$H = -\sum_{\langle ij \rangle} e^{-i A_{ij}} c_i^{\dagger} c_j + \text{h.c.}$. Our numerical methods and theoretical analysis hold even if we consider further neighbor hopping and/or interaction terms. 

\begin{figure*}[t]
    \centering
    \includegraphics[width=18cm]{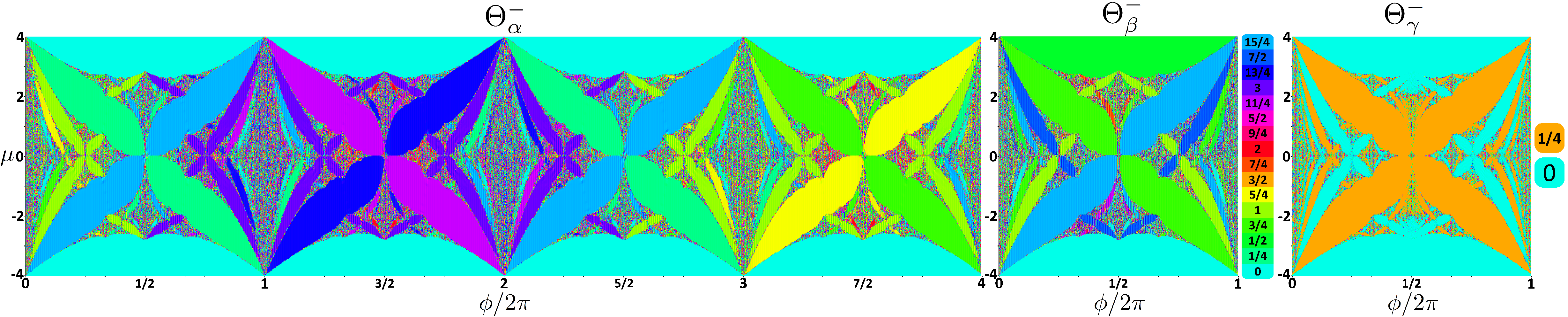}
    \caption{Raw data for $\Theta_{\OO}^-$ calculated on a $30\times 30$ open disc. The diameter of the partial rotation region $D$ is taken to be roughly half the system size. In the $C=\{1,2,3\}$ main Landau levels, $\Theta_{\beta}^{-}$ quantizes to $\{15/4\pm 0.003,
    1\pm 0.022, 3/4\pm 0.051\}$. $\Theta_{\OO}^-$ calculated on either an open disk or a torus yields the same result up to an exponentially small correction.}
    \label{fig:all}
\end{figure*}

\textit{Partial rotation calculation.} 
Consider a state $|\Psi\rangle$, which can be defined on a system with either open or periodic boundary conditions. 
Let $\Cmopm|_D$ be the restriction of $\Cmopm$ to an invariant subregion $D$ centered at $\OO$.

We find that 
$\bra{\Psi}\tilde{C}_{M_{\OO},\chi}^{\pm}|_D\ket{\Psi} =
e^{\frac{2\pi i}{M_{\OO}}l^\pm_{D,\OO,\chi} -\gamma^{\pm}_{D,\OO, \chi}}$. $\gamma^{\pm}_{D,\OO,\chi} \propto |\partial D|$ sets the real-valued amplitude. When $\chi$ is an integer, we find
\begin{equation}\label{eq:angular_momentum}
    l^\pm_{D,\OO,\chi}=\frac{C\chi^2}{2}+\mathscr{S}^{\pm}_{\OO}\chi + K^{\pm}_{\OO} \mod \MO
\end{equation}
$\mathscr{S}^+_{\OO}$ is the discrete shift, which was studied in Refs.\cite{zhang2022fractional,zhang2022pol} and was referred to there as $\mathscr{S}_{\OO}$. Here we define $\mathscr{S}^-_{\OO}$ analogously, as the term linear in $\chi$ in Eq. \ref{eq:angular_momentum}. 
Empirically we observe that $K^{\pm}_{\OO} \mod \MO$ may not be invariant throughout a given lobe in the Hofstadter model: $K^+_{\OO}$ can jump by multiples of $\MO/2$ within a single lobe, but there are no jumps in $K^-_{\OO}$. For an intuitive explanation of these jumps, which are system size dependent, see App.~\ref{app:num}.
We thus define
\begin{align}
    \Theta_{\OO}^+ :=
        K^+_{\OO} \mod \frac{\MO}{2}, \;\;\;\; \Theta_{\OO}^- := K^-_{\OO} \mod \MO.
\end{align}
The main result of this paper is that $\{\Theta_{\OO}^{\pm}\}$ are topological invariants protected by point group rotational symmetry about $\OO$.
$\Theta_{\OO}^+$ is in general an integer or half-integer, while $\Theta_{\OO}^-$ is a multiple of $1/4$. For a fixed Chern number, they can take $\MO/2$ and $2\MO$ different values respectively and thus define $\Z_{\MO/2}$ and $\Z_{2\MO}$ invariants, respectively. 
We find both empirically and analytically using Eq. \ref{eq:mainresultCFT} that $\Theta_\gamma^+ = 0 \mod 1$, and so it will henceforth be ignored. When $C = 0$, $\{\Theta_{\OO}^{\pm}\}$ are closely related to the many-body real-space indicators defined in \cite{herzogarbeitman2022interacting}.

The values of $\Theta_{\OO}^{\pm}$ are plotted for the square lattice Hofstadter model in Figs.~\ref{fig:beta_plus}, \ref{fig:alpha_plus}, and \ref{fig:all}. 
The values of $\Theta_{\OO}^+$ satisfy the empirical relation
\begin{equation}\label{eq:Theta+_vs_k3}
    \Theta_{\OO}^+ =
    k_{3,\OO}^+-\frac{C}{2} \mod 2, \;\; \OO = \alpha,\beta 
\end{equation}
where $k^+_{3,\OO}$ is an integer. The subscript 3 is used to match a convention from \cite{manjunath2021cgt}.
$\Theta_{\OO}^-$ satisfies
\begin{equation}\label{eq:Theta-_vs_k3}
    \Theta_{\OO}^- = \begin{cases}k^-_{3,\OO} + \frac{k_{s,\OO}}{2}-\frac{3C}{4} \mod 4, &  \OO = \alpha,\beta \\
    k_{3,\OO}^-+ \frac{k_{s,\OO}}{2} -\frac{C}{4} \mod 2, & \OO = \gamma,\end{cases}
\end{equation}
where $k_{s,\OO} \in \{0,1\}$ and $k_{3,\OO}^-$ is an integer.  Eqs.~\eqref{eq:Theta+_vs_k3},~\eqref{eq:Theta-_vs_k3} are in excellent agreement with a conformal field theory calculation, which predicts the basic quantization and $C$ dependence of $\Theta^{\pm}_{\OO}$ exactly, but does not give the actual values of $k_{3,\OO}^{\pm}$ and $k_{s,\OO}$ for a particular model. The CFT prediction is summarized by Eq.~\eqref{eq:mainresultCFT}.

If we view $\Theta^{\pm}_{\OO}$ in the square lattice Hofstadter model as a function of $\phi$, $\Theta^{\pm}_{\beta}$ has period $2\pi$ while $\Theta^{\pm}_{\alpha}$ has period $8\pi$. A similar periodicity depending on origin was observed for the discrete shift in Ref.~\cite{zhang2022pol}. Mathematically, the lack of periodicity in $2\pi$ arises because $\tilde{C}_{M_{\OO}}$ is not a gauge-invariant operator, and $H$ is only periodic (without a gauge transformation) under $\phi \rightarrow \phi + 8 \pi$ (For details see App.~\ref{app:num}). The lack of $2\pi$ periodicity is physically meaningful and arises because $\phi$ must be defined as a real number to consistently specify generic perturbations of $H$ \cite{zhang2022pol}.

The continuum limit of the Hofstadter model is obtained by taking $\phi \rightarrow 0^+, \nu \rightarrow 0^+$. In this limit states with Chern number $C$ are equivalent to $C$ filled Landau levels (LLs), and we have
\begin{align}
    \Theta^+_{\OO,\text{LL}} &= 
        \frac{C^3}{6}-\frac{2C}{3}\mod 2, \;\; \OO = \alpha,\beta
   \label{eq:Theta+LL}
   \\ 
    \Theta^-_{\OO,\text{LL}} &=
    \begin{cases}
        \frac{C^3}{6}+\frac{C^2}{4}-\frac{2C}{3}\mod 4& \OO = \alpha,\beta \\ \frac{C^3}{6}+\frac{C^2}{4}-\frac{C}{6}\mod 2 & \OO = \gamma
    \end{cases}
    . \label{eq:Theta-LL} 
\end{align}
As explained below, we can derive Eqs. \ref{eq:Theta+LL}, \ref{eq:Theta-LL}  analytically using results from CFT and TQFT, and also verify them empirically in our numerical calculations. 
$\Theta^{\pm}_{\OO}$ are closely related to the topological  term $\frac{\ell_{s,\OO}^{\pm}}{4\pi} \omega \wedge d \omega$ which appears in the effective response theory of continuum LLs coupled to an $\text{SO}(2)$ spin connection $\omega$ \cite{Wen1992shift}.

The empirical data for $\Theta^+_{\OO}$ can be fit to precise formulas throughout the square lattice Hofstadter model, which we use to fully color the butterflies in Fig.~\ref{fig:beta_plus},\ref{fig:alpha_plus}. These equations all have the general form
\begin{equation}\label{eq:Theta_full}
    \Theta^{\pm}_{\OO} = \Theta^{\pm}_{\OO,\text{LL}} + \Theta^{\pm}_{\OO,\text{diff}}.
\end{equation}
Eq.~\eqref{eq:Theta_full} indicates that we can express any given state with $C \ne 0$ as a stack of $C$ filled LLs (which contribute $\Theta^{\pm}_{\OO,\text{LL}}$) and a state with zero Chern number (which contributes $\Theta^{\pm}_{\OO,\text{diff}}$). Thus, $\Theta^{\pm}_{\OO,\text{diff}}$ specifies how the given state differs from the limit of $C$ filled LLs.
The empirical formulas for $\Theta^{+}_{\OO,\text{diff}}$ throughout the butterfly are summarized in App.~\ref{app:num}. We have not obtained empirical formulas for $\Theta^-_{\OO,\text{diff}}$ as these are substantially more complicated.

\paragraph*{Verification using conformal field theory.}
Eqs.~\eqref{eq:Theta+_vs_k3},~\eqref{eq:Theta-_vs_k3} can be analytically derived by using the cut-and-glue approach established in~\cite{Qi2012entanglement}, which describes the entanglement spectrum of the disk subregion in the long wavelength limit by that of the (1+1)D CFT on its edge. That is, the reduced density matrix for the disk subregion $D$ is effectively given by $\rho_{D} =\rho_{\mathrm{CFT}}$,
where $\rho_{\mathrm{CFT}}$ denotes the CFT on the edge of the disk.
The edge of the disk entangled with the complement subsystem is described by a thermal density matrix of a perturbed edge CFT within a fixed topological sector~\cite{Haldane2008entanglement}.
The form of the perturbation in the entanglement Hamiltonian is not universal.
In the following, we assume that the entanglement Hamiltonian is that of the unperturbed CFT: $\rho_{\mathrm{CFT}} = e^{-\beta H}$, and check the validity of this assumption with our numerics.

In accordance with the crystalline equivalence principle \cite{Thorngren2018,manjunath2022mzm}, the $\tilde C_{\MO}$ rotation symmetry acts as a translation symmetry combined with an internal $\Z_{\MO}$ symmetry of the CFT on the boundary of $D$. The expectation value of $\Cmopm$ for the disk $D$ without any additional flux insertion is hence evaluated in terms of the symmetry generators in the CFT as follows:
\begin{equation}
\begin{split}
&\bra{\Psi}\Cmopm|_D\ket{\Psi} = \frac{\mathrm{Tr}[e^{iQ_{\MO}\frac{\pi}{\MO}}e^{i\tilde{P}\frac{L}{\MO}}e^{-\frac{\xi}{v} H}]}{\mathrm{Tr}[e^{-\frac{\xi}{v} H}]} \\
&= e^{-\frac{2\pi i}{24\MO}c_-}\frac{\sum_{a=1,\psi}\chi_a(\frac{i\xi}{L}-\frac{1}{\MO};[\mathrm{AP},0],[\mathrm{AP},1])}{\sum_{a=1,\psi}\chi_a(\frac{i\xi}{L};[\mathrm{AP},0],[\mathrm{AP},0])}
\end{split}
\label{eq:partialrot_as_CFT}
\end{equation}
where we introduced the velocity $v$ of the CFT, finite temperature correlation length of the edge theory $\xi:=\beta v$, the length of the boundary $L=|\partial D|$, and $\{\text{AP,P}\}$ denotes the boundary condition with respect to $\Z_2^f$ fermion parity symmetry. The action of the $\MO$-fold rotation on the CFT is expressed as the combination $e^{iQ_{\MO}\frac{\pi}{\MO}}e^{i\tilde{P}\frac{L}{\MO}}$, where $Q_{\MO}$ generates an internal $\Z_{2\MO}^f$ (resp.~$\Z_{\MO}$) symmetry when we take the rotation symmetry to be $\Cmop$ (resp.~$\Cmom$). $\tilde{P}$ is the normalized translation operator 
\begin{align}
    \tilde{P}:=\frac{1}{v}(H-E_0) = \frac{2\pi}{L}\left[L_0-\frac{c_-}{24}-\langle L_0-\frac{c_-}{24}\rangle\right]
\end{align}
so that $\tilde{P}\ket{\mathrm{vac}}=0$ on the vacuum state $\ket{\mathrm{vac}}$ of the CFT.
$\chi_a(\tau;[s,j],[s',j'])$ with $s,s'\in\{\mathrm{AP},\mathrm{P}\}$, $j,j'\in\Z_{\MO}$ is the CFT character that corresponds to the partition function on a torus equipped with spin structure and $\Z_{\MO}$ gauge field.
As shown in App.~\ref{app:CFT}, the above CFT characters at high temperature $\xi \ll L$ can be evaluated using the modular $S,T$ matrices of the $G$-crossed braided fusion category \cite{barkeshli2019,manjunath2020FQH} describing the invertible phase. Our calculation uses the precise defect sectors of the $G$-crossed modular $S,T$ matrices that correspond to the boundary conditions which define the CFT character; the use of $G$-crossed modularity allows us to readily treat any symmetry action on the state encoded in the classification of fermionic invertible phases in (2+1)D~\cite{barkeshli2021invertible,aasen2021characterization}, some of whose details are missed in Ref.~\cite{shiozaki2017invt}. 
The results for even $\MO$ are to leading order given by 
\begin{align}
\bra{\Psi}\Cmop|_D\ket{\Psi}_{\text{CFT}} &\propto 
e^{-\frac{2\pi i}{24}(\MO-\frac{1}{\MO})c_-} \mathcal{I}_{M_{\OO}}^{+} = e^{\frac{2\pi i}{\MO}\Theta_{\OO}^+} \nonumber \\
\bra{\Psi}\Cmom|_D\ket{\Psi}_{\text{CFT}} &\propto e^{-\frac{2\pi i}{24}(\MO+\frac{2}{\MO})c_-} \mathcal{I}_{M_{\OO}}^- = e^{\frac{2\pi i}{\MO} \Theta_{\OO}^-}
\label{eq:mainresultCFT}
\end{align}
where 
\begin{align}
\label{IMeq}
\mathcal{I}_{M_{\OO}}^{\pm} := e^{\frac{2\pi i}{M_{\OO}} \frac{\ell_{s,\OO}^{\pm}}{2}} = e^{\frac{2\pi i}{\MO} \left( (1 \mp 1)\frac{ k_{s,\OO}}{4} + (1 \pm 1) \frac{c_-}{16} + k_{3,\OO}^{\pm} \right)} 
\end{align}
for integers $k^{\pm}_{3,\OO} \in \Z_{\MO}$ and $k_{s,\OO} \in \Z_2$. Eqs. \eqref{eq:mainresultCFT} and \eqref{IMeq} directly give Eqs.~\eqref{eq:Theta+_vs_k3},~\eqref{eq:Theta-_vs_k3}. 
Moreover, we have an equivalence $k^{+}_{3,\OO} \simeq k^{+}_{3,\OO} + \MO/2$ which comes from relabelling symmetry defects in the CFT by fermions, but no such equivalence for $k^{-}_{3,\OO}$.
We set $c_- = C$ to describe the non-interacting Hofstadter model. Combining Eqs.~\eqref{eq:mainresultCFT} and \eqref{IMeq} gives

\begin{align}\label{eq:ells_vs_Theta}
    \ell^{\pm}_{s,\OO} &=\begin{cases}
        \frac{11 \mp 1}{8}C+2\Theta^{\pm}_{\OO} \mod 4 \quad \OO=\alpha,\beta\\
        \frac{1\mp1}{4}C+2\Theta^{\pm}_{\OO} \mod 2\quad \OO=\gamma .
    \end{cases}
\end{align}

\paragraph*{Relation to topological action.}
$\ell^{\pm}_{s,\OO}$ appear in a general topological action which was derived for (2+1)D bosonic topological phases with symmetry $G = \text{U}(1) \times_\phi [\mathbb{Z}^2 \rtimes \mathbb{Z}_M]$ in Refs.~\cite{manjunath2021cgt,manjunath2020FQH}, and extended to invertible fermionic systems in the appendix of Ref.~\cite{zhang2022fractional}. It is written in terms of a $\text{U}(1)$ gauge field $A$, and crystalline gauge fields $(\vec{R},\omega)$. $\omega$ is a background `rotation' gauge field, which is treated as a real field with quantized holonomies. We can define $\omega$ as a $\Z_M$ gauge field or a $\Z_{2M}$ gauge field corresponding to the subgroups of $G$ generated by $\Cmop$ and $\Cmom$ respectively. We use $+$ and $-$ superscripts for the coefficients that appear in the action in each case. The coefficients with supercript $-$ can be obtained by replacing $A \rightarrow A + \omega/2$ in the action with only $+$ superscripts. 

The full action includes a term $\frac{\ell_{s,\OO}^{\pm} - c_-/12}{4\pi} \omega \wedge d\omega$, where
$\ell^{\pm}_{s,\OO}$ is quantized mod $\MO$, while $ - \frac{c_-}{12}$ is a contribution from the framing anomaly \cite{witten1989,Gromov2015}. The quantities $\ell^{\pm}_{s,\OO}$ determine the invariant of invertible phases protected purely by rotations about $\OO$ \cite{cheng2018rotation}, and arise in the CFT computation above through the $G$-crossed modular $S, T$ matrices. 
$\ell^{\pm}_{s,\OO}$ are directly related to $\Theta^{\pm}_{\OO}$, by Eq.~\eqref{eq:ells_vs_Theta}.
We defer a discussion of the remaining topological response coefficients, which include an angular momentum polarization $\vec{\mathscr{P}}_{s,\OO}^{\pm}$ and angular momentum filling $\nu_s^{\pm}$, to Ref. \cite{manjunath2023classif}.

For continuum Landau levels, $\ell^{\pm}_{s,\text{LL}}$ is an origin- independent $\Z$ invariant associated to continuous $\text{SO}(2)$ rotational symmetry. Now $\ell^{+}_{s,\text{LL}}$ is the coefficient of the term $\frac{1}{4\pi} \omega \wedge d\omega$ that arises in the effective action upon integrating out the fermion fields \cite{Wen1992shift}, $\mathcal{L} = \frac{1}{4\pi} \sum_{n=1}^C(A + s_n \omega) \wedge d(A + s_n \omega)$, where $s_n = \frac{2n-1}{2}$ is the spin of the fermion in the $n$th LL; and $\ell^{-}_{s,\text{LL}}$ is similarly obtained after replacing $A \rightarrow A + \omega/2$. Moreover, in the continuum limit of the Hofstadter model, we have
$\ell^{\pm}_{s,\OO,\text{LL}} = \ell^{\pm}_{s,\text{LL}} \mod \MO$. After simplification (see App.~\ref{app:LL}), we obtain 
\begin{align}\label{eq:ls-LL}
    \ell^+_{s,\OO,\text{LL}} &= \frac{C^3}{3}-\frac{C}{12} \mod \MO \nonumber \\
    \ell^-_{s,\OO,\text{LL}} &= \frac{C^3}{3}+\frac{C^2}{2}+\frac{C}{3} \mod \MO.
\end{align}
Combining Eq.~\eqref{eq:ls-LL} and Eq.~\eqref{eq:ells_vs_Theta} provides the analytical derivation of Eqs.~\eqref{eq:Theta+LL} and~\eqref{eq:Theta-LL}.

\paragraph*{Classification.} There are various relationships involving $\{\Theta^{\pm}_{\OO}\}$:
\begin{align}
\label{eq:Theta+-}
    2(\Theta^-_{\OO}-\Theta^+_{\OO}) &= \mathscr{S}^+_{\OO} \mod \MO,
\\
\Theta_\alpha^+ + \Theta_\beta^+ &=\Theta_\alpha^- + \Theta_\beta^- + 2 \Theta_\gamma^- - \frac{\kappa}{2} + C  \mod 2. \label{eq:nus+-}
\end{align}
Eqs.~\eqref{eq:Theta+-},~\eqref{eq:nus+-} can be derived in the LL limit (where $\kappa = 0$ but $C \ne 0$) using Eq. \eqref{eq:Theta+LL},~\eqref{eq:Theta-LL} and in the case of $C = 0$ using a real-space Wannier function argument \cite{manjunath2023classif}; combining the two results using linearity under stacking gives ~\eqref{eq:Theta+-},~\eqref{eq:nus+-}. Eq.~\eqref{eq:nus+-} implies that the 5 invariants $\{\Theta^{+}_{\alpha},\Theta^{+}_{\beta},\Theta^{-}_{\alpha},\Theta^{-}_{\beta},\Theta^{-}_{\gamma}\}$ define 4 independent invariants.

Eqs.~\eqref{eq:Theta+_vs_k3},~\eqref{eq:Theta-_vs_k3} imply that $I_1 := 2\Theta_\alpha^- + C/2$, 
$I_2 := \Theta_\alpha^+ + C/2$, and $I_3 := 2 \Theta_\gamma^- + C/2$ are integers modulo $8$, $2$, and $4$, respectively. Additionally,
 $I_4 := \Theta_\alpha^- + \Theta_\beta^- + 2\Theta_\gamma^- - \kappa/2$
is an integer modulo $4$. This can also be established in the LL limit using~\eqref{eq:Theta+LL},~\eqref{eq:Theta-LL} and in the case of $C = 0$ using a real-space Wannier function argument, and finally by combining the two using linearity under stacking. 

The above results imply that $\{I_1, I_2, I_3, I_4\} \in \Z_8 \times \Z_2 \times \Z_4 \times \mathbb{Z}_4$. But note that the overall classification of invertible fermionic phases with $G = \text{U}(1) \times_\phi [\mathbb{Z}^2 \rtimes \mathbb{Z}_4]$ is a group $\Z^3 \times \Z_8 \times \Z_2 \times\Z_4^2$ \cite{zhang2020realspace,manjunath2023classif}. The three $\mathbb{Z}$ invariants are $c_-$, $C$, $\kappa = \nu - \frac{C \phi}{2\pi}$. A subset of this classification which assumes $c_-=0, C=0$ was derived in Ref.~\cite{zhang2020realspace} through a real space construction. The full derivation will be explained in detail in a forthcoming work \cite{manjunath2023classif}, which also shows that $I_1, I_2, I_3, I_4$ are independent. From this, we can conclude that
$\{c_-, C, \kappa, I_1, I_2, I_3, I_4\}$, or equivalently $\{c_-, C, \kappa, \Theta^{\pm}_{\OO}\}$ fully characterize invertible fermionic states with symmetry group $G$.

\paragraph*{Discussion.} We have obtained the remarkable result that we can extract a \textit{complete} set of invariants (apart from $C,c_-,\nu$) from partial rotations, without inserting any extra magnetic flux or lattice defects. As a corollary, we can extract the previously studied invariants $\mathscr{S}^+_{\OO}, \vec{\mathscr{P}}^+_{\OO}$ in terms of $\{\Theta^{\pm}_{\OO},C,c_-,\nu\}$ as well, using Eq.~\eqref{eq:Theta+-} and the relationships between $\mathscr{S}^+_{\OO}$ and $\vec{\mathscr{P}}^+_{\OO}$ derived in \cite{zhang2022pol}. We also showed how to understand the topological response theory coefficients $\ell^{\pm}_{s,\OO}$ in terms of $\Theta^{\pm}_{\OO}$. 

We can instead study the eigenvalues of a \it global \rm rotation operator on a torus; this was used in \cite{zhang2022fractional,zhang2022pol} to find $\mathscr{S}^+_{\OO},\vec{\mathscr{P}}^+_{\OO}$, and also discussed in Ref.~\cite{herzogarbeitman2022interacting}. Some preliminary results for global rotations are shown in App.~\ref{app:global}; it is unclear how much of the classification can be obtained in this way.

It is worth highlighting that we have extracted $\{\Theta^{\pm}_{\OO}\}$ from a single wave function without requiring the insertion of additional magnetic flux or lattice defects. Note that one can also extract from a single wave function the quantities $c_-$ \cite{Haldane2008entanglement,FQHEDMRG,Qi2012momentumpolarization,Kim2022ccc}, $C$ \cite{shiozaki2017invt,dehghani2021,cian2021,fan2022}, and $\nu$. The above discussion then implies that one can in principle extract all the many-body invariants characterizing an invertible state with $G = \text{U}(1)\times_{\phi}[\Z^2\rtimes \Z_M]$ from a single wavefunction. 

Above, we focused on the square lattice, with $M=4$ and $M_{\OO} = 2$ or 4. For any orientation preserving space group symmetry, we have the following results \cite{manjunath2023classif}. If $\MO = 2, 4, 6$, $\Theta^+_{\OO}$ defines a $\Z_{\MO/2}$ invariant, while $\Theta^-_{\OO}$ defines a $\Z_{2\MO}$ invariant. (This means that if we know $C$, $\Theta^{+}_{\OO}$ and $\Theta^-_{\OO}$ can take $\MO/2$ and $M$ values respectively, although their precise quantization will depend on $C$.) On the other hand, if $\MO = 3$, both $\Theta^+_{\OO}$ and $\Theta^-_{\OO}$ define $\Z_{\MO}$ invariants. In either case, $\{\Theta^{\pm}_{\OO}\}$ along with $C, \nu$ and $c_-$, fully characterize the many-body crystalline invariants of systems with symmetry $\text{U}(1) \times_{\phi} [\mathbb{Z}^2 \rtimes \Z_M]$, and these can be extracted using a single wave function.

\paragraph*{Acknowledgements.}
This work is supported by NSF CAREER grant (DMR- 1753240), and the Laboratory for Physical Sciences through the Condensed Matter Theory Center. RK is supported by a JQI postdoctoral fellowship at the University of Maryland.

\bibliography{bibliography}
\clearpage
\appendix

\begin{widetext}

\section{Further details on the numerics}\label{app:num}

\subsection{Definition of vector potential}
In this section we provide a definition of the vector potential $A$ for the square lattice Hofstadter model on a torus. Some of the following details have previously appeared in Ref.~\cite{zhang2022fractional}. We first define our vector potential on a $L\times L$ torus with $L$ even. Considering even $L$ is sufficient for our calculations on the torus. An example of the gauge choice is shown in Fig.~\ref{fig:vector_potential}. We can use a similar gauge to define $A$ on an open disc. 

\begin{figure}[t]
    \centering
    \includegraphics[width=6.5cm]{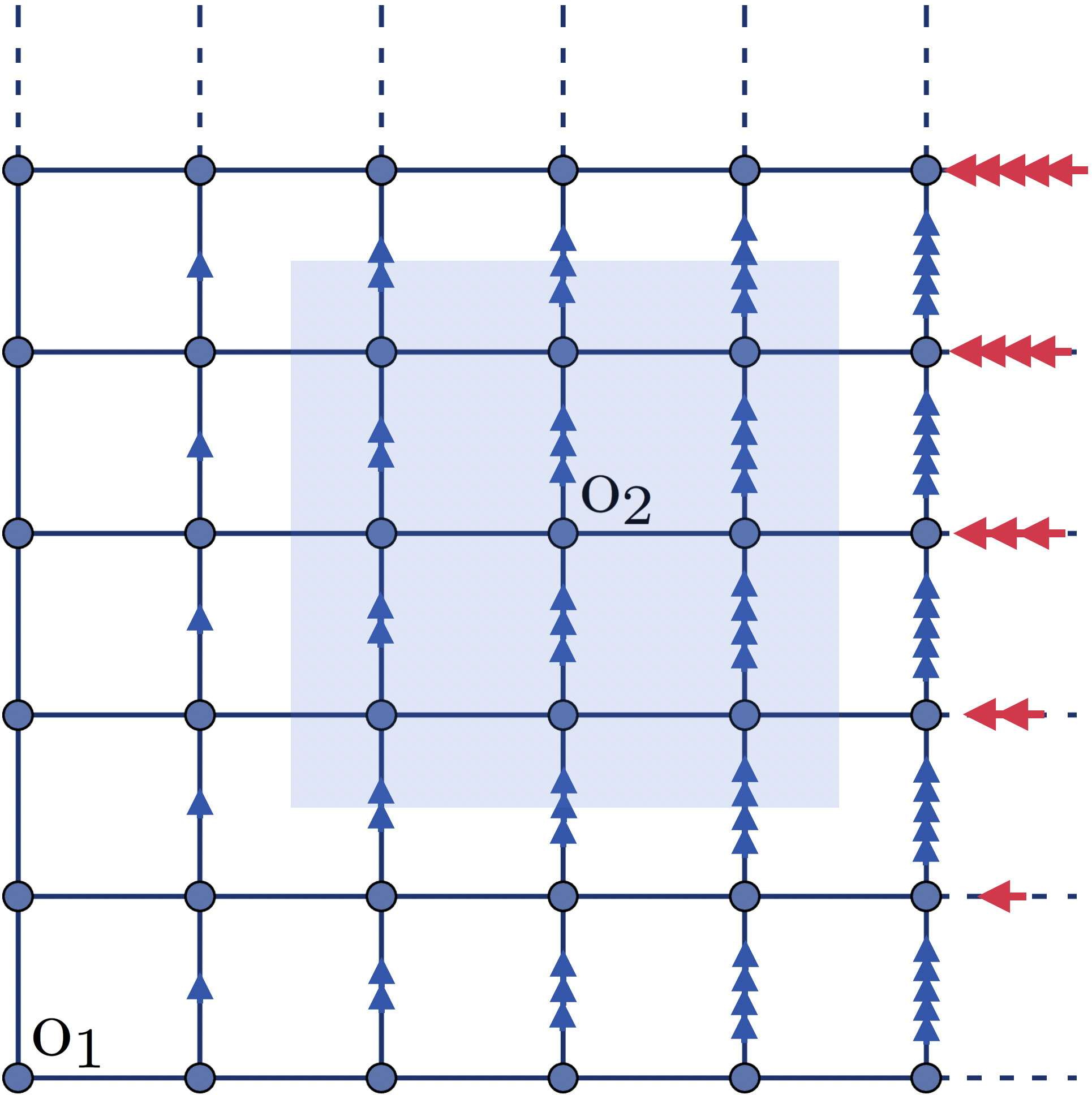}
    \caption{$L\times L=6\times 6$ square lattice with periodic boundary condition indicated by the dotted bonds. We insert $2\pi m$ flux uniformly through the torus, where $m = \frac{\phi}{2\pi} L^2$. Each blue arrow represents a hopping phase $A_{ij}=\frac{2\pi m}{L^2}$; each red arrow represents a hopping phase $A_{ij}=\frac{2\pi m}{L}$. $\OO_1,\OO_2$ are the rotation centers. The $3\times 3$ partial rotation region is defined around $\OO_2$ and is colored blue.}
    \label{fig:vector_potential}
\end{figure}

\subsection{Choice of global rotation operator}
In systems with $\text{U}(1)$ and point group rotation symmetries, we have the ambiguity that any rotation symmetry operator can be multiplied by a phase $e^{i \theta \hat{N}}$ to give another valid symmetry operator. It is necessary to fix this ambiguity in order to uniquely define a rotation operator. Below we illustrate how to do this on the infinite plane, and then on the torus.

\subsubsection{On the infinite plane}

Ref.~\cite{zhang2022fractional} showed how to canonically define one such operator ($\Cmop$ in the notation of this paper) on the infinite plane, which assigns zero excess flux around a disclination of angle $2\pi/\MO$ created at $\OO$ using $\Cmop$. Since the defect Hamiltonian depends on the specific choice of $\text{U}(1)$ gauge transformation defining $\Cmop$, considering a different operator $e^{i \chi \hat{N}} \Cmop$ modifies the flux around the disclination by $\chi \mod 2\pi$, and therefore we can physically distinguish it from $\Cmop$. See Ref.~\cite{zhang2022fractional} for the explicit construction of a disclination Hamiltonian with fourfold rotational symmetry.

Let us specialize to the case $\MO=4$ and $(\tilde{C}^+_{\MO})^4 = +1$. Assume each plaquette on the infinite plane has flux $\phi \mod 2\pi$.
If $\OO = \beta$ (a vertex), the four possible operators assign equal flux $\phi \mod 2\pi$ to all plaquettes in the lattice with a $\pi/2$ disclination, except one that lies next to the disclination. 
In this plaquette the assigned flux is $\phi + k\frac{\pi}{2} \mod 2\pi$, where $k = 0,1,2,3$. Thus there is a canonical choice of $\tilde{C}^+_{M_{\beta}}$, corresponding to $k=0$ (meaning zero excess flux). When $\OO = \alpha$ (a plaquette center), there is a triangular plaquette at a $\pi/2$ disclination which is assigned a flux $\frac{3\phi}{4} + k\frac{\pi}{2} \mod 2\pi$. The operator $\tilde{C}^+_{M_{\alpha}}$ for which $k=0$ is also canonical, because it assigns a flux at the disclination that is proportional to $\phi$.

To extract the desired topological invariants in this paper, it turns out to be insufficient to use the operators $\Cmop$ alone, because they only generate groups of order $\MO$. Thus we also define $\Cmom := e^{i \frac{\pi}{\MO} \hat{N}} \Cmop$, which is of order $2\MO$ and satisfies $(\Cmom)^{\MO} = (-1)^F$. The two sets of operators, together denoted $\Cmopm$, are indeed sufficient for our purpose. Arbitrary $\text{U}(1)$ rotations can be considered by defining $\tilde{C}_{\MO,\chi}^{\pm} = e^{i \chi \frac{2\pi}{\MO} \hat{N}} \Cmopm$. Note that $\tilde{C}_{\MO,\chi}^{+},\tilde{C}_{\MO,\chi}^{-}$ insert excess flux $2\pi \chi/\MO$ at a $2\pi/\MO$ disclination, compared to $\Cmop$ and $\Cmom$ respectively.

\subsubsection{On the torus}
Defining a unique rotation operator on a torus with fourfold rotational symmetry is slightly more subtle than on the infinite plane, because any order 4 rotation preserves 2 points $\OO_1,\OO_2$. If the side length $L$ of the torus is even, the two fixed points are either both at $\alpha$ or both at $\beta$. If $L$ is odd, we have one fixed point each at $\alpha$ and $\beta$. 

$\Cmop$ can be defined for $\OO = \OO_1$ or $\OO_2$ by restricting to a disc $D$ centered around $\OO$, and creating a $\pi/2$ disclination within $D$ centered at $\OO$, using $\Cmop|_D$. As in the case of the infinite plane discussed above, we canonically define $\Cmop$ such that $\Cmop|_D$ creates a disclination with zero excess flux. We then define $\Cmom := e^{i \frac{\pi}{\MO} \hat{N}}\Cmop$ and $\tilde{C}_{\MO,\chi}^{\pm} = e^{i \chi \frac{2\pi}{\MO} \hat{N}} \Cmopm$.

\subsection{Partial rotation on a torus}

Given the global rotation operators $\tilde{C}_{M_{\OO},\chi}^{\pm}$, we wish to restrict them to a region $D$, hence obtaining $\tilde{C}_{M_{\OO},\chi}^{\pm}|_D$. 
Note that the chosen operator inserts excess flux $2\pi\chi/\MO$ in a $2\pi/\MO$ disclination centered at $\OO$ relative to $\Cmopm|_D$. We find that the result for $l^{\pm}_{D,\OO,\chi}$ follows Eq.~\eqref{eq:angular_momentum}.

\subsection{Partial rotation on an open disc}

We can also perform the partial rotation calculation on an open disc. This can be done by first defining the vector potential similarly as in Fig.~\ref{fig:vector_potential}, using a gauge which inserts $\phi$ flux per plaquette in some region. As shown in Fig.~\ref{fig:general_setup}, It is \textit{not} necessary that the vector potential has support on the whole disc: the flux needs to be inserted only in an invariant subregion around $\OO$. Besides, $D$ \textit{does not} have to be defined around the center of the subregion with flux, as long as $D$ is enclosed by this subregion. The center of $D$, which we define as $\OO$, can be at any maximal Wyckoff point.

On the torus, we have a global rotation symmetry operator $\Cmopm$, and we define $\Cmopm|_D$ to be the restriction of $\Cmopm$ to region $D$. On the disc, this cannot be done as there might not be a global symmetry operator to start with. Therefore, we first choose $D$ and then define an operator $\Cmopm|_D$, which commutes with the Hamiltonian restricted to $D$. We find that choosing linear size $L_D$ of $D$ to be too large ($L_D \ge L-1$) or too small
($L_D \le 2$) gives unquantized $K^{\pm}_{\OO}$, which is tested this for $L \le 30$. As before,
we need to impose a constraint on $\tilde{C}_{4,\OO}|_D$ to fix the ambiguity by a global $\text{U}(1)$ phase $e^{i \theta \hat{N}}$. This can be done by fixing the excess flux inserted around a disclination created using the $\Cmopm|_D$ operator as explained in the previous section. We define $\tilde{C}_{\MO,\chi}^{+}$ (respectively $\tilde{C}_{\MO,\chi}^{-}$) so that it inserts an excess flux of $2\pi\chi/\MO$ at a $2\pi/\MO$ disclination, relative to $\Cmop, \Cmom$.

Note that, we obtain the same result, Eq.~\eqref{eq:angular_momentum}, on the torus or the open disc, as long as we choose $\OO$ at the same maximal Wyckoff position, and choose a partial rotation operator that inserts the same flux around a disclination at $\OO$.

\begin{figure}[t]
    \centering
    \includegraphics[width=8cm]{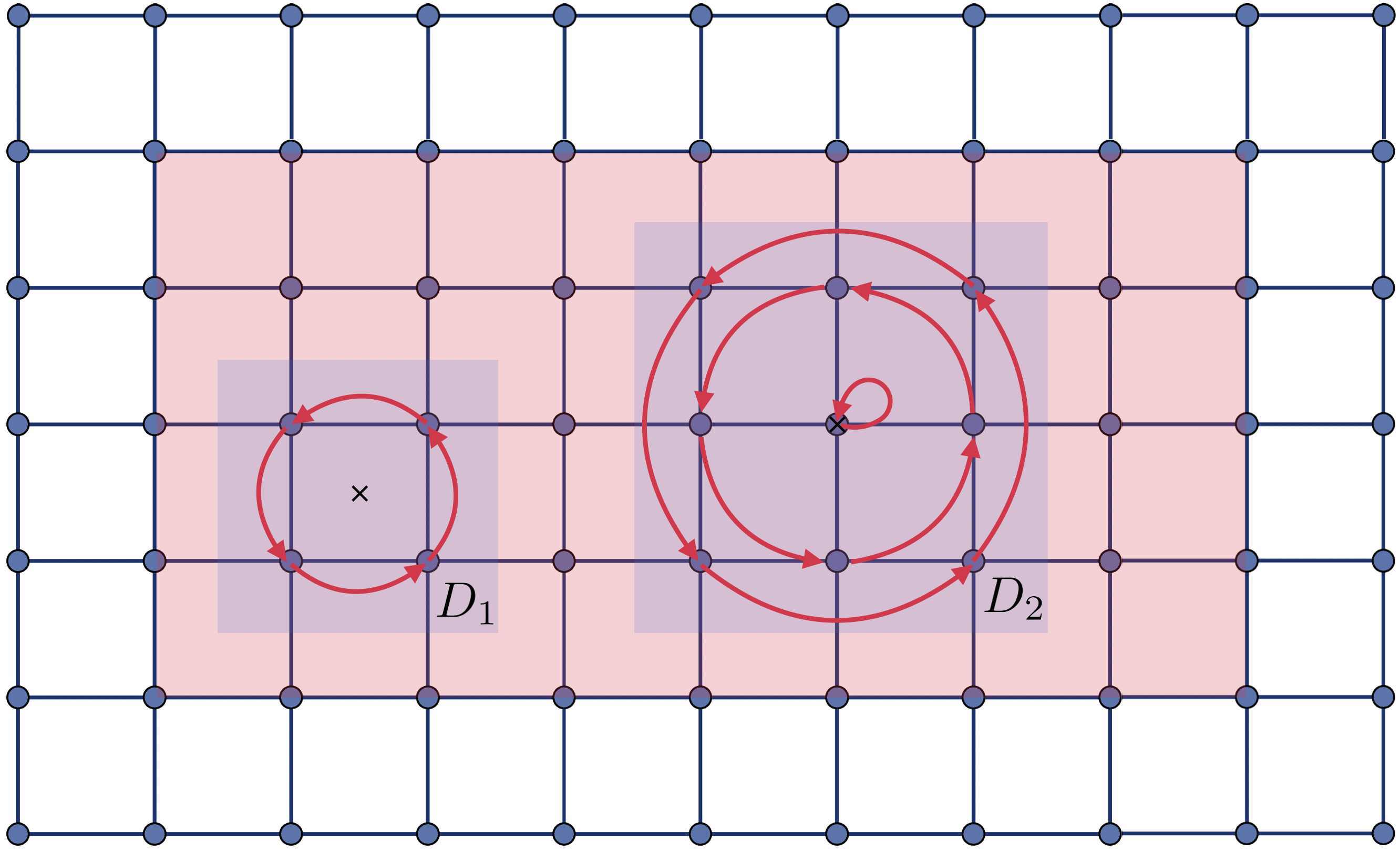}
    \caption{Calculation on an open disc. Red shading indicates the sub-region where the flux is inserted. Blue shading inside the red region indicates different choices of maximal Wyckoff point as the partial rotation center.}
    \label{fig:general_setup}
\end{figure}

\subsection{Details on the jumps of $K_{\OO}^+$}

\begin{figure}[t]
    \centering
    \includegraphics[width=7cm]{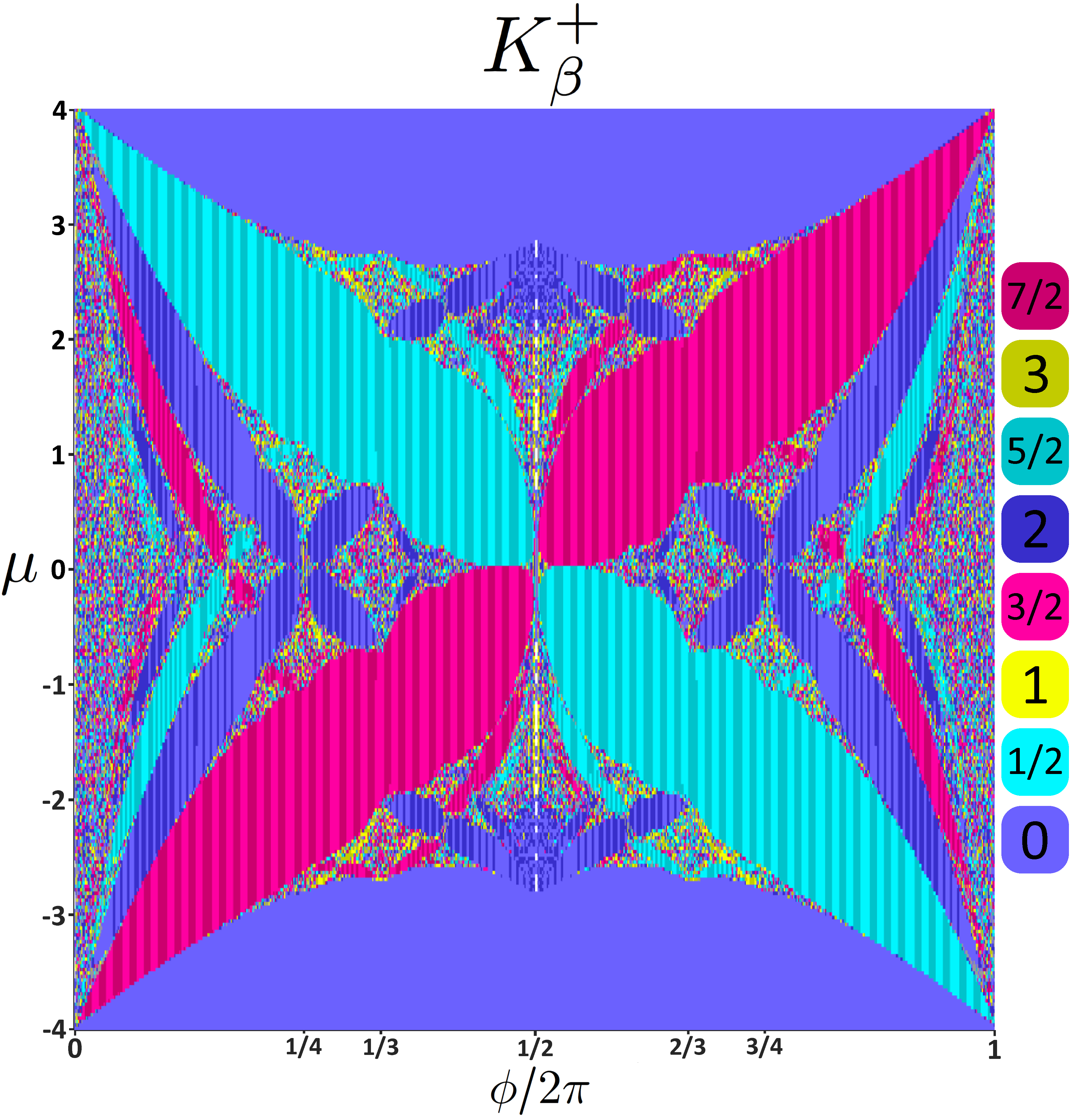}
    \caption{Raw data for $K_\beta^{+}$, calculated on a $30\times 30$ open disc. Partial rotation region $D$ is a $15\times 15$ square. $K_{\beta}^{+}$ jumps by 2 within a single lobe whenever the color switches between bright and dim shades. }
    \label{fig:jumps}
\end{figure}

As stated in the main text, if we use the operators $\Cmop$, $K^+_{\OO}$ sometimes jumps by $\MO/2$ inside a single Hofstadter lobe, as seen in Fig.~\ref{fig:jumps}. We define $\Theta_{\OO}^{+}\equiv K^+_{\OO} \mod \MO/2$ in order to eliminate these jumps. In general, the position of these jumps will depend on the system size and the size of the partial rotation region, but the value of these jumps is always $\MO/2$. We can understand these jumps as follows. The following arguments are equivalent to those given in \cite{herzogarbeitman2022interacting} using the many-body real space invariants defined there.

Suppose $\MO=4$ in a system where all orbitals are filled ($\nu = 1$). First consider $K^+_{\OO}$. Let $K_{\OO}^{+}|_{\nu=1}$ be the value of $K^+_{\OO}$ at $\nu=1$. Now suppose we minimally enlarge the partial rotation region $D$ by enclosing $4$ more sites which rotate into each other Under $\Cmop$. The states at these 4 sites form an orbit of size 4 under $\Cmop$ and contribute extra phases of $\{1, e^{i2\pi/4},e^{i4\pi/4},e^{i6\pi/4}\}$, which are fourth roots of 1. The expectation value of $\Cmop$ will gain a total extra phase of $e^{i\pi}$ (which is the product of the 4 phases), and thus $K_{\OO}^{+}|_{\nu=1}$ changes by 2 mod 4. But a change of the size of the partial rotation region should be regarded as trivial. Therefore only $K^+_{\OO} \mod 2$ is an invariant quantity. 

On the other hand, considering $\Cmom$ in the same argument, the states at the 4 extra sites in the enlarged region contribute phases of $\{e^{i\pi/4}, e^{i3\pi/4},e^{i5\pi/4},e^{i7\pi/4}\}$, which are fourth roots of -1. The total phase contribution from these four sites is 1, and thus $K_{\OO}^-$ is invariant upon changing $D$. 

Since changing $D$ is trivial, the above example suggests that $\Theta_{\OO}^+$ is an invariant mod $\MO/2$, and $\Theta_{\OO}^-$ is an invariant mod $\MO$. In the latter case we indeed verify numerically that there are no jumps of $K^-_{\OO}$ inside a given Hofstadter lobe.

\subsection{Periodicities of $\Theta_{\OO}^{\pm}$}

\begin{figure}[t]
    \centering
    \includegraphics[width=8cm]{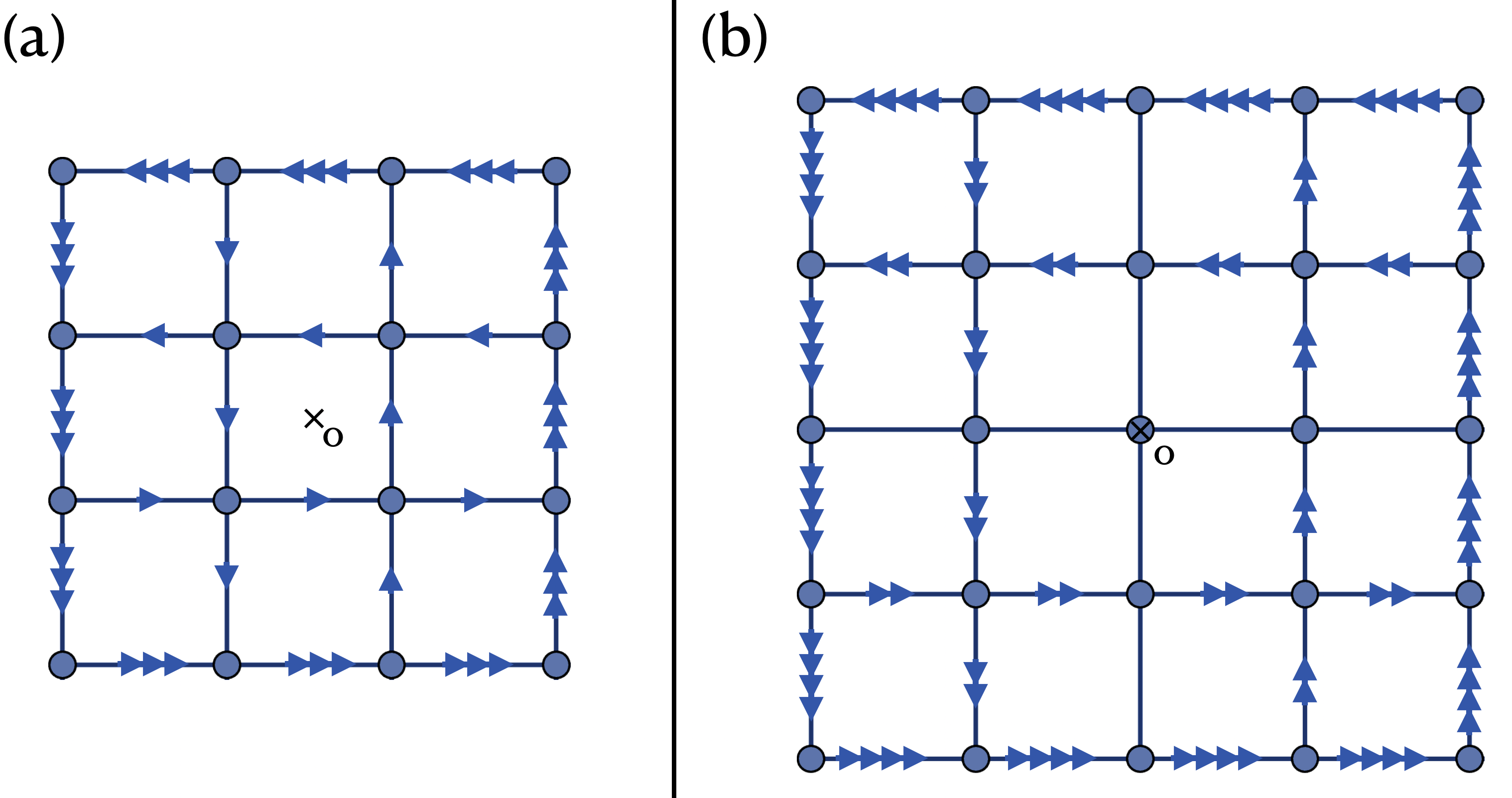}
    \caption{Symmetric gauge defined with respect to \textbf{(a)} $\alpha$; \textbf{(b)} $\beta$. Each blue arrow represents the vector potential $\phi/4$.}
    \label{fig:periodicity}
\end{figure}

We find that as a function of the background flux per unit cell in the Hofstadter model, $\Theta^{\pm}_{\beta}$ has periodicity $2\pi$, while $\Theta^{\pm}_{\alpha}$ has periodicity $8\pi$. We can straightforwardly see why these invariants need not be $2\pi$ periodic in $\phi$. Consider a system on the infinite plane, with flux $\phi$ per unit cell. For a given origin $\OO$, it is most convenient to define the vector potential in symmetric gauge around $\OO$; in this case the operator that trivially rotates points without any $\text{U}(1)$ gauge transformation is also a symmetry operator that commutes with the Hamiltonian.

As shown in Fig.~\ref{fig:periodicity}, if $\OO = \beta$, the links are assigned a vector potential which is an integer multiple of $\pi/2$, while if $\OO = \beta$, the links are assigned a vector potential which is an integer multiple of $\pi/4$. Therefore $H$ is $4\pi$ periodic in $\phi$ when $\OO = \beta$, and $8\pi$ periodic when $\OO = \alpha$. The actual periodicity of $\Theta^{\pm}_{\beta},\Theta^{\pm}_{\alpha}$ in $\phi$ must therefore be divisible by $4\pi$ and $8\pi$ respectively. The fact that $\Theta^{\pm}_{\beta}$ has a smaller periodicity $2\pi$ while $\Theta^{\pm}_{\alpha}$ has the maximal allowed periodicity of $8\pi$ is a more subtle result that we see empirically, but do not have a simple justification for on a lattice without defects. 

However, we can consider the flux inserted at a $\pi/2$ disclination created using $\tilde{C}_{M_{\beta}}^{\pm}$ and $\tilde{C}_{M_{\alpha}}^{\pm}$ respectively. As seen previously, for $\OO = \beta$ the flux is always expressed in terms of integer multiples of $\phi$ (which is $2\pi$ periodic), while for $\OO = \alpha$ the flux is expressed in terms of the quantity $3\phi/4$ (which is $8\pi$ periodic). Therefore defects of $\tilde{C}_{M_{\beta}}^{\pm}$ and $\tilde{C}_{M_{\alpha}}^{\pm}$ do have the same periodicities in $\phi$ as $\Theta^{\pm}_{\beta},\Theta^{\pm}_{\alpha}$. This gives additional justification for the observed periodicities.

\subsection{Obtaining empirical formulas for $\Theta_{\OO}^+$}

In this section, we explain how to obtain empirical formula for $\Theta_{\alpha}^+$ 
 and $\Theta_{\beta}^+$. The result is in Eqs.~\eqref{eq:theta_alpha}, \eqref{eq:theta_beta}.

In Fig.~\ref{fig:all} and ~\ref{fig:theta_plus_raw} we plot the raw Hofstadter butterflies for $\Theta_{\OO}^{\pm}$. For a fixed Chern number $C$, the different Hofstadter lobes are separated by the so called Farey sequence of order $2|C|$, which consists of ordered irreducible fractions $\frac{p}{q}$ with $0<p \le q\le 2|C|$. For example, the Farey sequence of order 4 is $\{\frac{1}{4},\frac{1}{3},\frac{1}{2},\frac{2}{3},\frac{3}{4},\frac{1}{1}\}$. If we track the lobes at a fixed $C$ as $\phi/2\pi$ is increased from 0 to 1, $\Theta_{\OO}^{\pm}$ may change its value between lobes which meet at the Farey seq of order $|C|$. (A similar behaviour was observed for the discrete shift $\mathscr{S}_{\OO}^+$ in Ref.~\cite{zhang2022fractional}.)

\begin{figure*}[t]
    \centering
    \includegraphics[width=17.5cm]{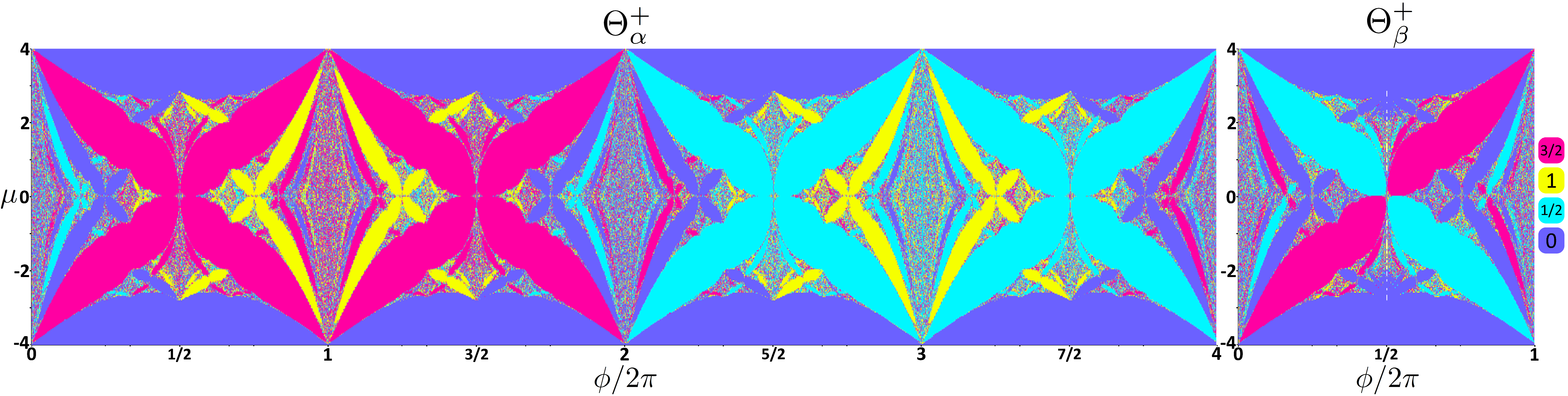}
    \caption{Raw data for $\Theta_{\OO}$. $\Theta_\alpha^+$, $\Theta_\beta^+$, and $\Theta_{\gamma}^+$ are calculated on a $30\times 30$ open disc. $\Theta_{\gamma}^+=0 \mod 1$ everywhere in the Hofstadter butterfly, and is therefore not plotted here. The diameter of the partial rotation region $D$ is taken to be roughly half the system size. $\Theta_{\OO}^+$ calculated on either an open disk or a torus yields the same result up to an exponentially small correction.}
    \label{fig:theta_plus_raw}
\end{figure*}
 
The jumps for $\Theta_{\alpha}^+$ and $\Theta_{\beta}^+$ are tabulated in Figs.~\ref{fig:fareya}, \ref{fig:fareyb} and can be categorized into two contributions: A jump of 1 at every odd denominator $q$ for $q<|C|$; and a \textit{possible} jump of 1 when $q$ divides $C$. When $C>0$ we find that after summing the contribution from each jump point,
\begin{align}
    \Theta^+_{\alpha,\text{diff}} &=   \left(\sum_{\substack{\frac{p}{q}<\frac{\phi}{2\pi}\\\text{odd }q
    }} \left\lfloor \frac{C+q}{2q} \right\rfloor \right) + \begin{cases}\left\lfloor\frac{C\phi}{2\pi}\right\rfloor \text{if } C \mod 4 =2\\
    \left\lfloor\frac{C\phi}{4\pi}\right\rfloor \text{if } C \mod 4 =3\\
    0 \text{ if } C \mod 4 =0\\ 
    \left\lfloor\frac{1}{2}+\frac{C\phi}{4\pi}\right\rfloor \text{if } C \mod 4 =1
    \end{cases}\label{eq:theta_alpha}\\
    \Theta^+_{\beta,\text{diff}} &=  \left(\sum_{\substack{\frac{p}{q}<\frac{\phi}{2\pi}\\\text{odd }q
    }} \left\lfloor \frac{C+q}{2q} \right\rfloor \right)  + \begin{cases}\left\lfloor\frac{C\phi}{2\pi}\right\rfloor \text{if } C \mod 4 =1\\
    \left\lfloor\frac{C\phi}{4\pi}\right\rfloor \text{if } C \mod 4 =2\\
    0 \text{ if } C \mod 4 =3\\
    \left\lfloor\frac{1}{2}+\frac{C\phi}{4\pi}\right\rfloor \text{if } C \mod 4 =0
    \end{cases}\label{eq:theta_beta}\\
    \Theta^+_{\gamma,\text{diff}} &= 0.
\end{align}
These equations are all taken mod $\MO/2$. $\Theta^+_{\alpha}$ in the $C<0$ lobes can be obtained by the symmetry transformation $\Theta^+_{\alpha}(\mu,\phi)=\Theta^+_{\alpha}(-\mu,\phi) \mod 2$; $\Theta^+_{\beta}$ in the $C<0$ lobes can be obtained by the symmetry transformation $\Theta^+_{\beta}(\mu,\phi)=-\Theta^+_{\beta}(-\mu,\phi)\mod 2$. We have not found the analogous formulas for $\Theta^{-}_{\OO}$, as the jump patterns are different and much more complicated in this case.

\begin{figure}[t]
    \centering
    \includegraphics[width=8cm]{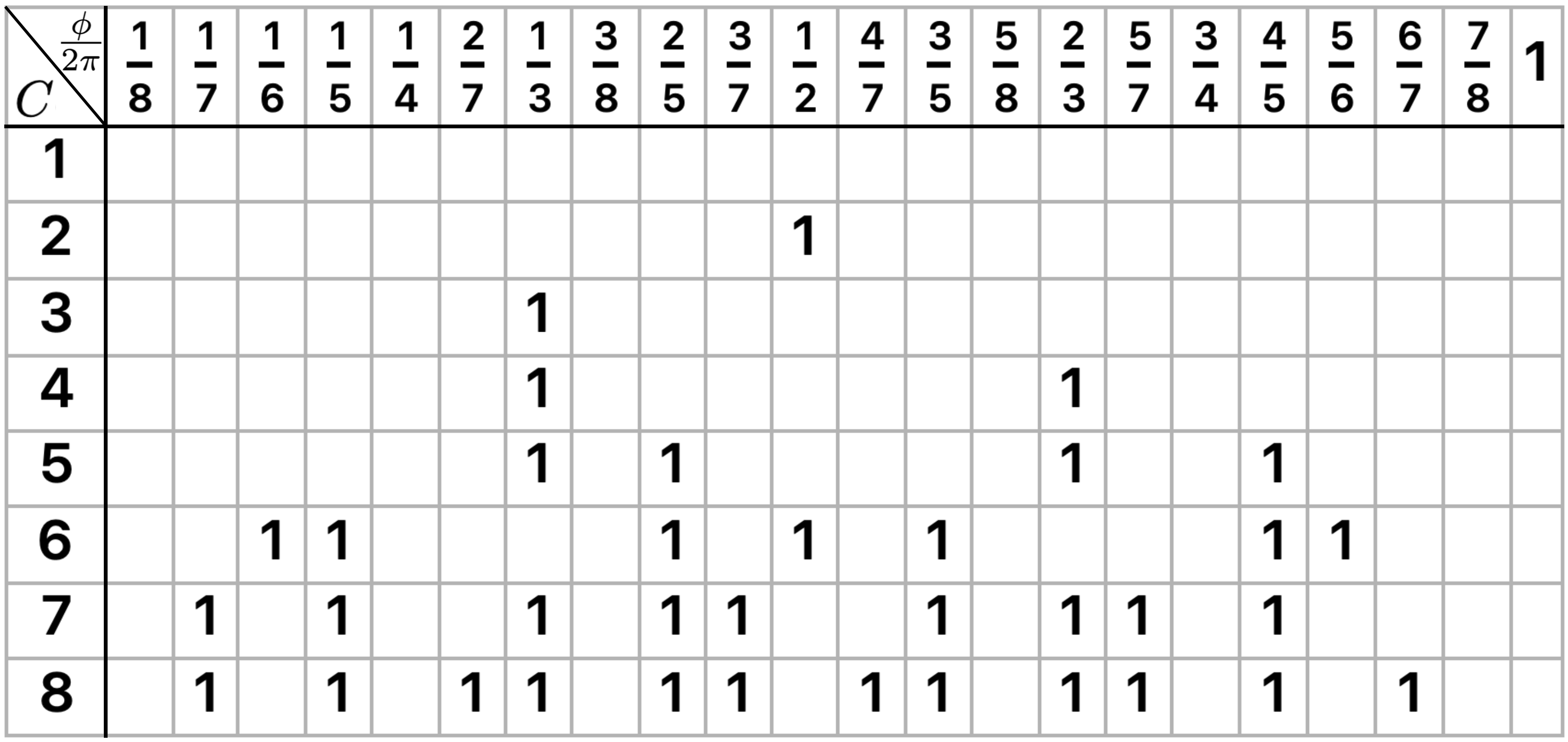}
    \caption{Jumps in $\Theta_{\alpha}^+$ for fixed $C$, as $\frac{\phi}{2\pi}$ increases from 0 to $\frac{1}{2}$.}
    \label{fig:fareya}
\end{figure}
\begin{figure}[t]
    \centering
    \includegraphics[width=8cm]{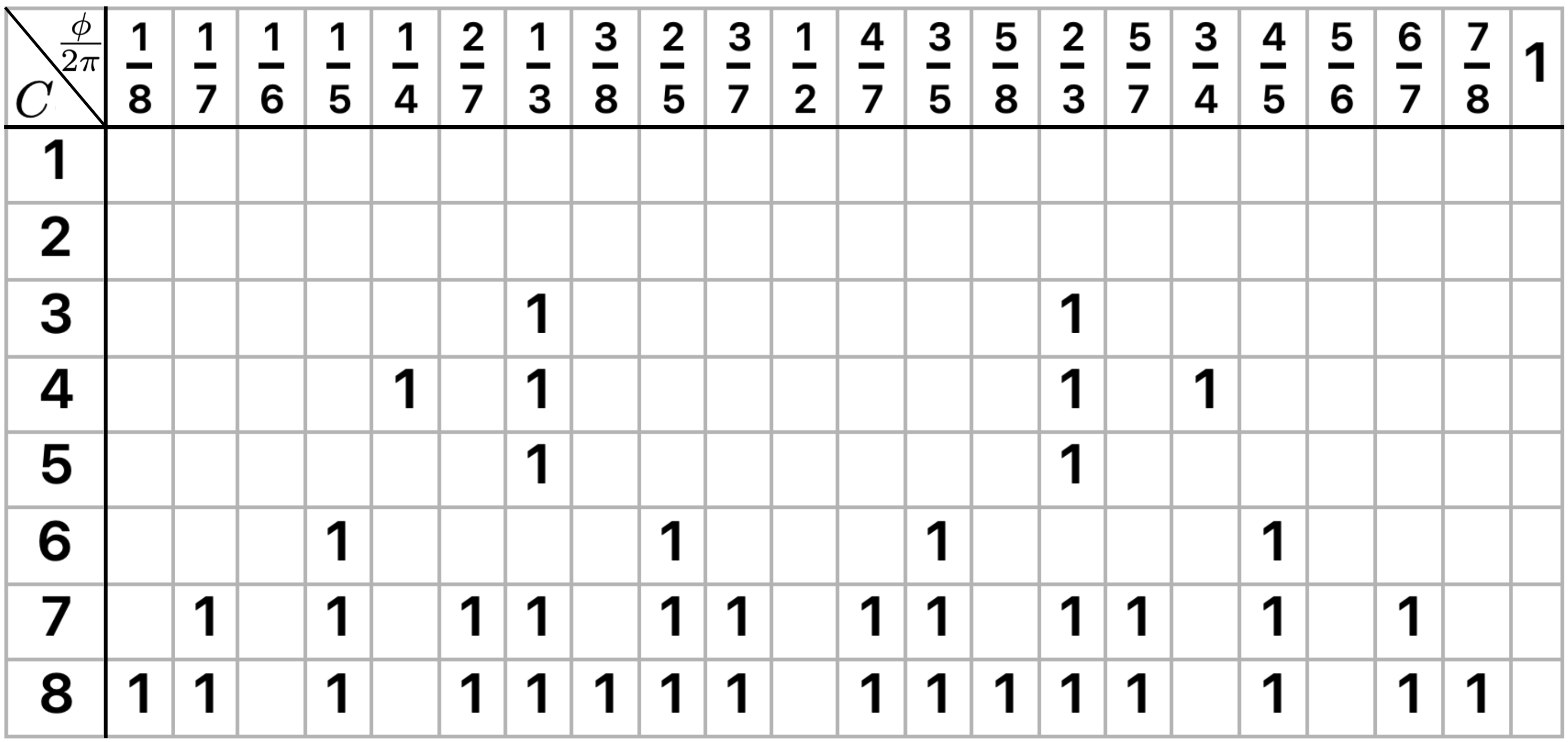}
    \caption{Jumps in $\Theta_{\beta}^+$ for fixed $C$, as $\frac{\phi}{2\pi}$ increases form 0 to $\frac{1}{2}$}
    \label{fig:fareyb}
\end{figure}

\section{Numerical results for global rotations}\label{app:global}

In this section we present some preliminary numerical results for the invariants of the Hofstadter model obtained from the eigenvalue under global rotations, $\bra{\Psi}\Cmopm\ket{\Psi}$, on the ground state $\ket{\Psi}$ on an even length torus. 

Define $\Cmopm$ as in the previous section. We define a set of invariants $\Phi_{\OO}^{\pm}$ as follows:
\begin{equation}
    \bra{\Psi}\Cmopm\ket{\Psi}  = e^{i l^{\pm}_{\OO,\text{global}}},
\end{equation}
and we empirically find that
\begin{align}
    l^{\pm}_{\OO_2,\text{global}} &= \frac{Cm^2}{2} + \mathscr{S}^{\pm}_{\OO_2}m + K_{\OO_2,\text{global}}^{\pm}\mod 4 \\
    l^{\pm}_{\OO_1,\text{global}} &= K_{\OO_1,\text{global}}^{\pm}\mod 4.
\end{align}

If we fix a given Hofstadter lobe, the quantities $K_{\OO,\text{global}}^+$ jump by multiples of $\MO/2$ as the total length of the torus is changed. (Recall that for partial rotations in a region $D$, the quantities $K_{\OO}^+$ similarly jump by multiples of $\MO/2$ for the same system, as the size of $D$ is changed.) However there is no change in the value of $K_{\OO,\text{global}}^-$. So we define
\begin{align}
    \Phi_{\OO}^{+} &= K^+_{\OO,\text{global}} \mod \MO/2 \nonumber \\
    \Phi_{\OO}^{-} &= K^-_{\OO,\text{global}} \mod \MO.
\end{align}
The raw Hofstadter butterflies for $\{\Phi_{\OO}^{\pm}\}$ are shown in Fig.~\ref{fig:fullrotation}. We see that for either $\OO = \OO_1$ or $\OO = \OO_2$,
\begin{align}
    \Phi^+_{\alpha} = \Phi^+_{\beta} = \Phi^+_{\gamma} &= \Phi^-_{\gamma} = -C \mod 2 \label{eq:cmod2} \\ 
    \Phi^-_{\alpha} = \Phi^-_{\beta} &= -C \mod 4. \label{eq:cmod4}
\end{align} Ref.~\cite{herzogarbeitman2022interacting} also studied global rotations on an even length torus in terms of a many-body real space invariant which appears closely related to $\Phi_{\OO}^{\pm}$, and also found that this invariant only depends on $C$. In the above cases, these invariants do not give any additional information beyond the Chern number. It is not clear how much more information can be obtained by also considering odd length tori. 

\begin{figure}[t]
    \centering
    \includegraphics[width=10cm]{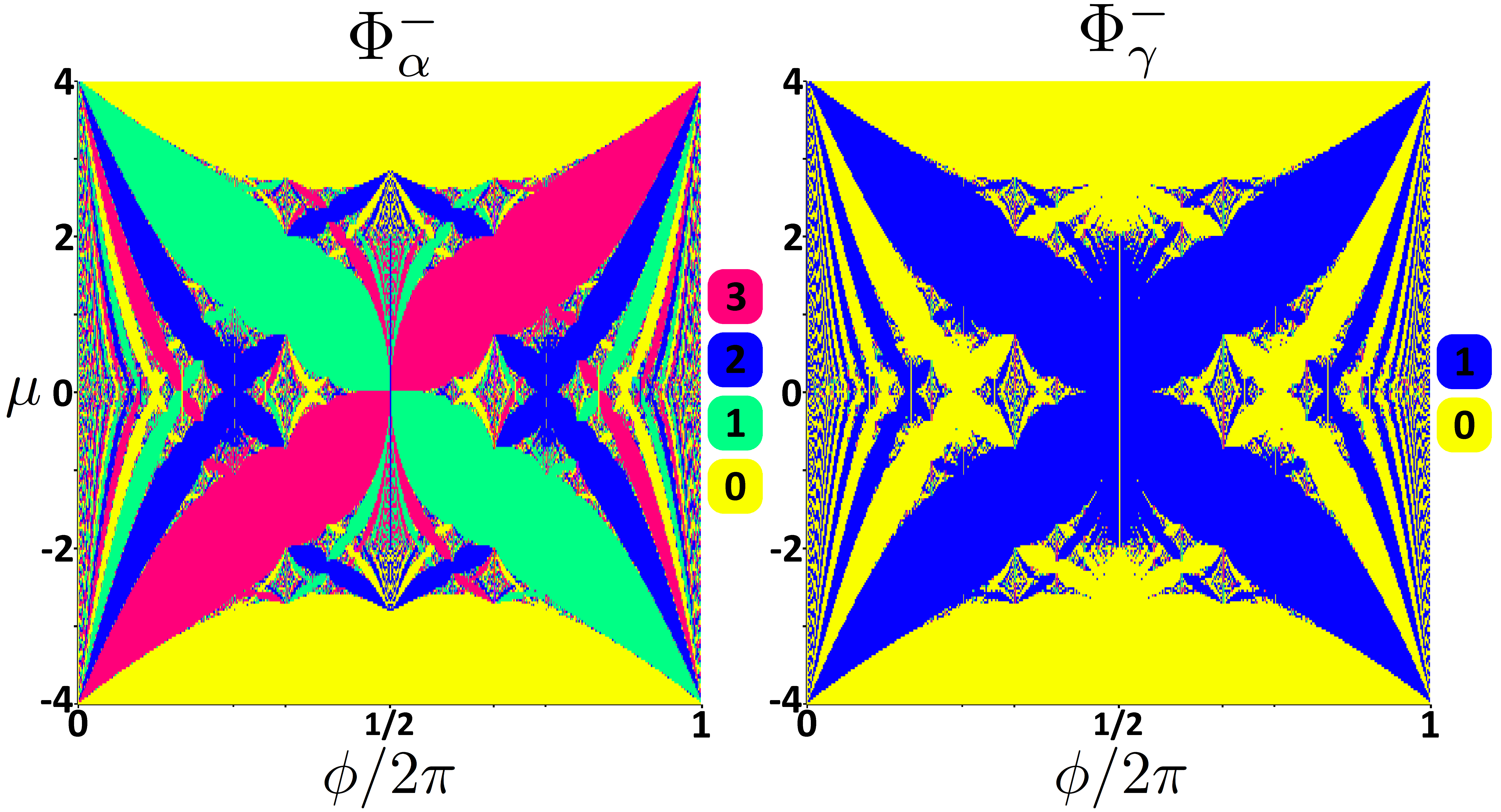}
    \caption{Raw data for $\Phi^-_{\alpha}$ and $\Phi^+_{\gamma}$, calculated on a $30\times 30$ torus. They follow Eq.~\ref{eq:cmod4} and Eq.~\ref{eq:cmod2} respectively.}
    \label{fig:fullrotation}
\end{figure}

\section{CFT calculation}
\label{app:CFT}
In this section, we derive the expression Eq.~(10) in the main text for the partial rotation by evaluating the CFT character,
\begin{equation}
\begin{split}
&\bra{\Psi}\tilde{C}^\pm_{M}|_D\ket{\Psi} = \frac{\mathrm{Tr}[e^{iQ_M\frac{\pi}{M}}e^{i\tilde{P}\frac{L}{M}}e^{-\frac{\xi}{v} H}]}{\mathrm{Tr}[e^{-\frac{\xi}{v} H}]} \\
&= e^{-\frac{2\pi i}{24M}c_-}\frac{\sum_{a=1,\psi}\chi_a(\frac{i\xi}{L}-\frac{1}{M};[\mathrm{AP},0],[\mathrm{AP},1])}{\sum_{a=1,\psi}\chi_a(\frac{i\xi}{L};[\mathrm{AP},0],[\mathrm{AP},0])}.
\label{eq:rotcharacter}
\end{split}
\end{equation}
For ease of notation, we will use $M$ in this section and suppress the $\OO$ subscript, but we assume that some $\OO$ has been fixed. While comparing with numerics we take $M = \MO=4$ for $\OO = \alpha, \beta$ and $M=2$ for $\OO = \gamma$.
Here, $\chi_a(\tau;[s,j],[s',j'])$ with $s,s'\in\{\mathrm{AP},\mathrm{P}\}$, $j,j'\in\Z_M$ is the CFT character that corresponds to the partition function on a torus equipped with a spin structure and a background $\Z_M$ gauge field. For example,
\begin{align}
    \chi_a(\tau;[\mathrm{AP},j],[\mathrm{AP},j'])=\mathrm{Tr}_{a,[\mathrm{AP},j]}[e^{iQ_M\frac{j'\pi}{M}}e^{2\pi i \tau (L_0-\frac{c_-}{24})}]
\end{align}
where $a$ labels the quasiparticle within the AP (antiperiodic) sector, with the twisted boundary condition corresponding to $j \in \Z_n$ along the spatial cycle. 

In order to evaluate the above CFT character, we need to study different cases of the action of the rotation operator $\tilde{C}^\pm_{M}$, i.e., whether $(\tilde{C}^+_{M})^M=+ 1$ or $(\tilde{C}^-_{M})^M=(-1)^F$. This choice of the symmetry action amounts to considering the twisted or untwisted spin structure respectively for the $\Z_M$ gauge field coupled to the CFT. That is, the rotation symmetry $\tilde{C}_{M}^{\pm}$ at long wavelengths is expressed as a combination of an internal $\Z_M$ symmetry and the translation symmetry of the CFT: $\tilde{C}_M^\pm=e^{iQ_M\frac{\pi}{M}}e^{i\tilde{P}\frac{L}{n}}$. Note that the symmetry action of $\tilde{C}_M^+$ (resp.~$\tilde{C}_M^-$) is realized by an internal symmetry satisfying $e^{iQ_M\pi}=(-1)^F$ (resp.~$e^{iQ_M\pi}=1$). The former case corresponds to a $\Z_{2M}^f$ symmetry where the $\Z_M$ is nontrivially extended by the fermion parity $(-1)^F$, which corresponds to the twisted spin structure, or equivalently a spin$^{\Z_{2M}}$ structure. Meanwhile, the latter corresponds to the untwisted spin structure together with a flat $\Z_M$ gauge field.

The above identification of the spatial symmetry $\tilde{C}_M^+   (\tilde{C}_M^-)$ with an internal symmetry $\Z_{2M}^f (\Z_M \times \Z_2^f)$ can be thought of a consequence of the crystalline equivalence principle \cite{Thorngren2018,manjunath2022mzm}. As we will see below, these two choices of the symmetry action give rise to different modular properties of the CFT characters, and hence distinct values of $\bra{\Psi}\tilde{C}^\pm_{M}|_D\ket{\Psi}$.

\subsection{The case with $\tilde{C}_M^+$}
Let us first take the rotation symmetry $\tilde{C}_M^+$, which has an equivalent internal $\Z_{2M}^f$ symmetry satisfying $e^{iQ_M\pi}=(-1)^F$. We perform the computation by cases of even or odd $M$.

\subsubsection{Even $M$}
For even $M$, the CFT character on the edge can be evaluated by the modular $S,T$ transformations as follows: 
\begin{align}
\begin{split}
\chi_a\left(\frac{i\xi}{L}-\frac{1}{M};[\mathrm{AP},0],[\mathrm{AP},1]\right)&= S_{ab}\chi_b\left(-\frac{1}{\frac{i\xi}{L}-\frac{1}{M}};[\mathrm{AP},1],[\mathrm{AP},0]\right) \\
&= (ST^M)_{ab}\chi_b\left(\frac{-iM\frac{\xi}{L}}{\frac{i\xi}{L}+\frac{1}{M}};[\mathrm{AP},1],[\mathrm{P},0]\right),
\end{split}
\end{align}
where we used the fact that $T$ exchanges Spin$^{\Z_{2M}}$ structure as follows:
\begin{align}
T:\quad
\begin{cases}
     ([\mathrm{AP},j],[\mathrm{AP},j']) & \to ([\mathrm{AP},j],[\mathrm{P}+\frac{[j]_M +[j']_M-[j+j']_M}{M},[j+j']_M]), \\
    ([\mathrm{AP},j],[\mathrm{P},j']) & \to ([\mathrm{AP},j],[\mathrm{AP}+\frac{[j]_M +[j']_M-[j+j']_M}{M},[j+j']_M]). \\
    \end{cases}
\end{align}
Here, $[]_{M}$ denotes the mod $M$ operation, and the spin structure changes under $\Z_2$ action as $\mathrm{AP}+1=\mathrm{P}$, $\mathrm{P}+1=\mathrm{AP}$.

To evaluate the action of the $S,T$ modular matrices in the presence of a background $\Z_{2M}^f$ gauge field, we note that the modular transformation in the defect Hilbert space of the CFT can be determined from the modular $S,T$ matrices of the $G_b$-crossed braided fusion category that describes the bosonic shadow of the invertible phase of interest with bosonic global symmetry $G_b$ in the bulk~\cite{barkeshli2019}. In our case, the bosonic symmetry group is $G_b = \Z_M$. Using the data of the $\Z_M$-crossed braided fusion category, the $T$-matrix element is expressed as~\cite{barkeshli2019}
\begin{align}
    T^{(\mathbf{g},\mathbf{h})}_{a_{\mathbf{g}},b_{\mathbf{g}}} = e^{-\frac{2\pi i}{24}c_-}\cdot \theta_{a_{\mathrm{g}}}\cdot \eta_a(\mathbf{g},\mathbf{h})\cdot\delta_{a_{\mathbf{g}},b_{\mathbf{g}}}
\end{align}
where $\mathbf{g},\mathbf{h}\in \Z_M$, and $\eta_a(\mathbf{g},\mathbf{h})$ is a phase that describes the symmetry fractionalization of the bulk topological order~\cite{barkeshli2019}.
The character is then further rewritten as
\begin{align}
   \begin{split}
&\chi_a\left(\frac{i\xi}{L}-\frac{1}{M};[\mathrm{AP},0],[\mathrm{AP},1]\right)\\
&= \sum_{b\in\mathcal{C}_{1}}(ST^M)_{ab}\chi_b\left(\frac{-iM\frac{\xi}{L}}{\frac{i\xi}{L}+\frac{1}{M}};[\mathrm{AP},1],[\mathrm{AP},0]\right) \\
&= e^{-\frac{2\pi iM}{24}c_-}\sum_{b\in\mathcal{C}_{1}} S_{ab} \theta_b^n \prod_{j=0}^{M-1} \eta_b(1,j)\times \chi_b\left(\frac{-iM\frac{\xi}{L}}{\frac{i\xi}{L}+\frac{1}{M}};[\mathrm{AP},1],[\mathrm{P},0]\right) \\
&= e^{-\frac{2\pi iM}{24}c_-}\sum_{b\in\mathcal{C}_{{1}}}\sum_{c\in\mathcal{C}_{\mathrm{P}}} S_{ab} \theta_b^M \prod_{j=0}^{M-1} \eta_b\left(1,j\right)\times S_{bc} \chi_c\left(\frac{iL}{M^2\xi}+\frac{1}{M};[\mathrm{P},0],[\mathrm{AP},1]\right),
\end{split} 
\end{align}
and
\begin{align}
    \begin{split}
       &\chi_a\left(\frac{i\xi}{L};[\mathrm{AP},0],[\mathrm{AP},0]\right)= \sum_{b\in\mathcal{C}_0}S_{ab} \chi_b\left(\frac{iL}{\xi};[\mathrm{AP},0],[\mathrm{AP},0]\right),
    \end{split}
    \label{eq:trivialcharacter}
\end{align}
where $\mathcal{C}_0$ (resp.~$\mathcal{C}_{1}$) is the untwisted (resp.~twisted by $1\in\Z_M$) sector in $G_b$-crossed extension of super-modular category that consists of defects in the $[\mathrm{AP},0]$ sector, namely the identity particle and the fermion (resp. the $[\mathrm{AP},1]$ defects, namely the elementary $\Z_M$ fluxes). Meanwhile, $\mathcal{C}_{\mathrm{P}}$ denotes the defects in the $[\mathrm{P},0]$ sector that are labelled by fermion parity.

The character in Eq.~\eqref{eq:trivialcharacter} can further be approximated as
\begin{align}
\begin{split}
    \chi_b\left(\frac{iL}{\xi};[\mathrm{AP},0],[\mathrm{AP},0]\right) &\approx e^{-\frac{2\pi L}{\xi}(h_b-\frac{c_-}{24})}~.
    \end{split}
    \label{eq:characterapproxfermion}
\end{align}
The phase of the partial rotation can then be written as
\begin{align}
\begin{split}
 \bra{\Psi}\tilde{C}_{M}^+|_D \ket{\Psi}  & \propto {e^{-2\pi i(M+\frac{1}{M})\frac{c_-}{24}}}\sum_{b\in\mathcal{C}_{1}}\sum_{c\in\mathcal{C}_{\mathrm{P}}} (S_{1b}+S_{\psi b}) \theta_b^n \prod_{j=0}^{M-1} \eta_b\left(1,j\right)\times S_{bc} \chi_c\left(\frac{iL}{M^2\xi}+\frac{1}{M};[\mathrm{P},0],[\mathrm{AP},1]\right).
   \end{split}
   \end{align}
   While the above expression requires the sum over quasiparticles in the twisted sector $b\in\mathcal{C}_1$, it is more illuminating to express it in terms of the anyons in the untwisted sector. This is done by using the consistency equations and Verlinde formula of the $\Z_M$-crossed braided fusion category, given by~\cite{barkeshli2019}
\begin{align}
\begin{split}
    \theta_{b_{1}} &= \theta_{b_0}\theta_{0_{1}}\cdot (R^{b_0, 0_{1}}R^{0_{1}, b_0}), \\
    \eta_{b_1}\left(1,j\right) &= \eta_{b_0}\left(1,j\right)\eta_{0_1}\left(1,j\right), \\
    S_{a,b_{1}} &= \frac{1}{d_a}S_{a,b_0}(R^{a, 0_{-1}}R^{0_{-1}, a}),
\end{split}
\end{align} 
where $0_j\in \mathcal{C}_j$ denotes a $j\in \Z_M$ defect.
From the general solution outlined in Ref.~\cite{barkeshli2019},
one can work in the gauge where $R^{b,0_{1}}=1$, $R^{c,0_1}=1$ for $b\in\mathcal{C}_0$, $c\in\mathcal{C}_{\mathrm{P}}$. We then obtain
\begin{align}
\begin{split}
    \bra{\Psi}\tilde{C}^+_{M}|_D \ket{\Psi}  \propto &e^{-\frac{2\pi i}{24}(M+\frac{1}{M})c_-} \theta_{0_{1}}^M\prod_{j=0}^{M-1}\eta_{0_1}\left(1,j\right)\sum_{b\in\mathcal{C}_0}\sum_{c\in\mathcal{C}_{\mathrm{P}}} d_b \theta_b^M \prod_{j=0}^{M-1} \eta_b\left(1,j\right)
   \times S_{bc} \chi_c\left(\frac{iL}{M^2\xi}+\frac{1}{M};[\mathrm{P},0],[\mathrm{AP},1]\right).
   \end{split}
   \end{align}
Here we observe invariants given by the following combination of symmetry fractionalization data:
\begin{align}
    \mathcal{I}_M^{\pm}:= \theta_{0_{1}}^M\prod_{j=0}^{M-1}\eta_{0_1}\left(1,j\right),
\end{align}
where the $\pm$ superscript is used depending on whether we consider $\Tilde{C}_M^+$ or $\tilde{C}_M^-$, and
\begin{align}
    e^{i\pi Q_b}:= \prod_{j=0}^{M-1} \eta_b\left(1,j\right).
\end{align}
They can be interpreted as a $\Z_M$ analog of the Hall conductivity, and a fractional $\Z_M$ charge respectively \cite{manjunath2020FQH}.
For $\frac{L}{\xi}\gg 1$, we approximate the CFT character in terms of the highest weight state $\ket{h_b}$: 
\begin{align}
\begin{split}
    \chi_c\left(\frac{iL}{M^2\xi}+\frac{1}{M};[\mathrm{P},0],[\mathrm{AP},1]\right) &\approx e^{\frac{2\pi i}{M}(h_c-\frac{c_-}{24})} e^{-\frac{2\pi L}{M^2\xi}(h_c-\frac{c_-}{24})}~.
    \end{split}
\end{align}
We then have
\begin{align}
\begin{split}
   \bra{\Psi}\tilde{C}^+_{M}|_D \ket{\Psi}   \propto &e^{-\frac{2\pi i}{24}(M+\frac{2}{M})c_-} e^{\frac{2\pi i}{M}h_v}\mathcal{I}^{+}_M\sum_{b\in\mathcal{C}_0} d_b\theta_b^M e^{i\pi Q_b} S_{bv},
   \end{split}
   \end{align}
where $v\in\mathcal{C}_{\mathrm{P}}$ is the quasiparticle in the periodic sector (fermion parity flux) with the lowest value of spin.
For the Chern insulator with chiral central charge $c_-$, there are two such defects $v, v \times \psi$ with equal topological spin $c_-/8$. The result with $\mathcal{C}_0=\{1,\psi\}$ is then given by
\begin{align}
\begin{split}
  \bra{\Psi}\tilde{C}^+_{M}|_D \ket{\Psi}  \propto &e^{-\frac{2\pi i}{24}(M-\frac{1}{M})c_-} \mathcal{I}^+_M.
   \end{split}
   \label{eq:partialrotresult_CMM1}
   \end{align}

   \subsubsection{Odd $M$}
For odd $M$, the CFT character is transformed by modular $S,T$ transformations as follows: 
\begin{align}
\begin{split}
\chi_a\left(\frac{i\xi}{L}-\frac{1}{M};[\mathrm{AP},0],[\mathrm{AP},1]\right)&= S_{ab}\chi_b\left(-\frac{1}{\frac{i\xi}{L}-\frac{1}{M}};[\mathrm{AP},1],[\mathrm{AP},0]\right) \\
&= (ST^M)_{ab}\chi_b\left(\frac{-iM\frac{\xi}{L}}{\frac{i\xi}{L}+\frac{1}{M}};[\mathrm{AP},1],[\mathrm{AP},0]\right).
\end{split}
\end{align}
Note that the temporal boundary condition in the last expression is shifted from P to AP compared with the case of even $M$.
Then, by using the same logic as the case of even $M$, one can obtain to the leading order
\begin{align}
\begin{split}
    \bra{\Psi}\tilde{C}^+_{M}|_D \ket{\Psi}  \propto &e^{-\frac{2\pi i}{24}(M+\frac{1}{M})c_-} \theta_{0_{1}}^M\prod_{j=0}^{M-1}\eta_{0_1}\left(1,j\right)\sum_{b\in\mathcal{C}_0}\sum_{c\in\mathcal{C}_{0}} d_b \theta_b^M \prod_{j=0}^{M-1} \eta_b\left(1,j\right)
   \times S_{bc} \chi_c\left(\frac{iL}{M^2\xi}+\frac{1}{M};[\mathrm{AP},0],[\mathrm{AP},1]\right) \\
   \propto & e^{-\frac{2\pi i}{24}(M+\frac{2}{M})c_-} \mathcal{I}^{+}_M\sum_{b\in\mathcal{C}_0} d^2_b\theta_b^M e^{i\pi Q_b}.
   \end{split}
   \end{align}
The result with $\mathcal{C}_0=\{1,\psi\}$ is then given by
\begin{align}
\begin{split}
  \bra{\Psi}\tilde{C}^+_{M}|_D \ket{\Psi}  \propto &e^{-\frac{2\pi i}{24}(M+\frac{2}{M})c_-} \mathcal{I}^+_M.
   \end{split}
   \end{align}

\subsection{The case with $\tilde C_M^-$}   
Here we consider the case where we take $\tilde C_M^-$ symmetry, which corresponds to the internal $\Z_M$ symmetry satisfying $e^{iQ_M\pi}=1$.

\subsubsection{Even $M$}
For even $M$, the CFT character on the edge can be evaluated by the modular $S,T$ transformation as
\begin{align}
\begin{split}
\chi_a\left(\frac{i\xi}{L}-\frac{1}{M};[\mathrm{AP},0],[\mathrm{AP},1]\right)&= S_{ab}\chi_b\left(-\frac{1}{\frac{i\xi}{L}-\frac{1}{M}};[\mathrm{AP},1],[\mathrm{AP},0]\right) \\
&= (ST^M)_{ab}\chi_b\left(\frac{-iM\frac{\xi}{L}}{\frac{i\xi}{L}+\frac{1}{M}};[\mathrm{AP},1],[\mathrm{AP},0]\right),
\end{split}
\end{align}
where we used $T$ exchanges spin$\times{\Z_{M}}$ structure as
\begin{align}
T:
\begin{cases}
    ([\mathrm{AP},j],[\mathrm{AP},j'])\to ([\mathrm{AP},j],[\mathrm{P},[j+j']_M]), \\
    ([\mathrm{AP},j],[\mathrm{P},j'])\to ([\mathrm{AP},j],[\mathrm{AP},[j+j']_{M}]). \\
    \end{cases}
\end{align}
Using a similar discussion as the previous subsection, one can write the phase of the partial rotation as
\begin{align}
\begin{split}
   \bra{\Psi}\tilde{C}^-_{M}|_D \ket{\Psi} & \propto {e^{-2\pi i\frac{c_-}{24} (M +\frac{1}{M})}}\mathcal{I}^-_M\sum_{b,c\in\mathcal{C}_0}d_b\theta_b^M \prod_{j=0}^{M-1} \eta_b\left(1,j\right)\times S_{bc} \chi_c\left(\frac{iL}{M^2\xi}+\frac{1}{M};[\mathrm{AP},0],[\mathrm{AP},1]\right),
   \end{split}
   \end{align}
   where we work on the gauge $R^{b,0_{\frac{\pi}{n}}}=1$ for $b\in\mathcal{C}_0$.
   The dominant contribution for the sum over anyons $c$ in the trivial sector comes from $c=1$, where the CFT character is approximated as
   \begin{align}
\begin{split}
    \chi_1\left(\frac{iL}{M^2\xi}+\frac{1}{M};[\mathrm{AP},0],[\mathrm{AP},1]\right) &\approx  e^{-\frac{2\pi i}{M}\frac{c_-}{24}} e^{\frac{2\pi L}{M^2\xi}\frac{c_-}{24}}~.
    \end{split}
\end{align}
We then obtain
\begin{align}
\begin{split}
  \bra{\Psi}\tilde{C}_{M}^-|_D \ket{\Psi} \propto &e^{-\frac{2\pi i}{24}(M+\frac{2}{M})c_-} \mathcal{I}^-_M\sum_{b\in\mathcal{C}_0} d^2_b\theta_b^M e^{i\pi Q_b}~.
   \end{split}
   \end{align}
For Chern insulators, in which $\mathcal{C}_0=\{1,\psi\}$, the partial rotation is simply given by
\begin{align}
\begin{split}
   \bra{\Psi}\tilde{C}_{M}^-|_D \ket{\Psi}  \propto &e^{-\frac{2\pi i}{24}(M+\frac{2}{M})c_-} \mathcal{I}^-_M.
   \end{split}
   \label{eq:partialrotresult_CMM(-1)F}   \end{align}

   \subsubsection{Odd $M$}
For odd $M$, the CFT character on the edge can be evaluated by the modular $S,T$ transformation as
\begin{align}
\begin{split}
\chi_a\left(\frac{i\xi}{L}-\frac{1}{M};[\mathrm{AP},0],[\mathrm{AP},1]\right)&= S_{ab}\chi_b\left(-\frac{1}{\frac{i\xi}{L}-\frac{1}{M}};[\mathrm{AP},1],[\mathrm{AP},0]\right) \\
&= (ST^M)_{ab}\chi_b\left(\frac{-iM\frac{\xi}{L}}{\frac{i\xi}{L}+\frac{1}{M}};[\mathrm{AP},1],[\mathrm{P},0]\right).
\end{split}
\end{align}
Note that the temporal boundary condition in the last expression is shifted from AP to P compared with the case of even $M$.
Then, by using the same logic as the case of even $M$, one can obtain to the leading order
\begin{align}
\begin{split}
    \bra{\Psi}\tilde{C}^+_{M}|_D \ket{\Psi}  \propto &e^{-\frac{2\pi i}{24}(M+\frac{1}{M})c_-} \theta_{0_{1}}^M\prod_{j=0}^{M-1}\eta_{0_1}\left(1,j\right)\sum_{b\in\mathcal{C}_0}\sum_{c\in\mathcal{C}_{\mathrm{P}}} d_b \theta_b^M \prod_{j=0}^{M-1} \eta_b\left(1,j\right)
   \times S_{bc} \chi_c\left(\frac{iL}{M^2\xi}+\frac{1}{M};[\mathrm{P},0],[\mathrm{AP},1]\right) \\
   \propto & e^{-\frac{2\pi i}{24}(M+\frac{2}{M})c_-}e^{\frac{2\pi i}{M}h_v}\mathcal{I}^{-}_M\sum_{b\in\mathcal{C}_0} d^2_b\theta_b^M e^{i\pi Q_b}S_{bv},
   \end{split}
   \end{align}
   where $v\in\mathcal{C}_{\mathrm{P}}$ is the quasiparticle in the periodic sector (fermion parity flux) with spin $c_-/8$.
   The result with $\mathcal{C}_0=\{1,\psi\}$ is then given by
\begin{align}
\begin{split}
  \bra{\Psi}\tilde{C}^+_{M}|_D \ket{\Psi}  \propto &e^{-\frac{2\pi i}{24}(M-\frac{1}{M})c_-} \mathcal{I}^-_M.
   \end{split}
   \end{align}

\subsection{Calculation of $\mathcal{I}^{\pm}_M$}

In Eq.~\eqref{eq:partialrotresult_CMM1},~\eqref{eq:partialrotresult_CMM(-1)F}, we observed that the partial rotation is proportional to the phase $\mathcal{I}^{\pm}_M$ that should be determined by $\tilde C^{\pm}_M$ symmetry action on the state. Here, we derive the general expression for $\mathcal{I}^{\pm}_M$ in terms of the data of a given fermionic invertible phase with $G_b = \Z_M$ symmetry and chiral central charge $c_-$.

The bosonic shadow of this phase is a 16-fold way $G_b$-enriched topological phase, also with chiral central charge $c_-$. In our setup, the symmetry does not permute anyons. Let the symmetry fractionalization data be given by
\begin{align}
    \omega_2({\bf g},{\bf h}) &= k_{\pm} \frac{[{\bf g}]_M+[{\bf h}]_M-[{\bf gh}]_M}{M} \mod 2 \\
    n_2({\bf g},{\bf h}) &= k_s \frac{[{\bf g}]_M+[{\bf h}]_M-[{\bf gh}]_M}{M} \mod 2.
\end{align}

$\omega_2$ describes the extension of the internal symmetry $\Z_M$ by fermion parity. From the fermionic crystalline equivalence principle, $k_+ = 1 \mod 2$ while $k_- = 0 \mod 2$. For a longer discussion of this point, see for example Ref.~\cite{manjunath2022mzm}. The two cases can be combined by defining $k_{\pm} = (1 \pm 1) \frac{1}{2}$.

If ${\bf g}_0$ is the generator of $G_b$, we wish to compute
\begin{equation}
     \mathcal{I}^{\pm}_M:= \theta_{0_{{\bf g}_0}}^M\prod_{j=0}^{M-1}\eta_{0_{{\bf g}_0}}\left({\bf g}_0,{\bf g}_0^j\right).
\end{equation}

We use the general solution outlined in Ref.~\cite{barkeshli2019} combined with the general theory of invertible fermionic phases in Ref.~\cite{barkeshli2021invertible}, which provides explicit expressions for the $\eta$ and $\theta$ symbols in terms of $n_2, \omega_2, c_-$. In particular, for this symmetry we can choose a gauge in which $ \theta_{0_{{\bf g}_0}} = 1$ and
\begin{equation}
    \eta_{0_{\bf k}}\left({\bf g},{\bf h}\right) = \nu_3^{-1}({\bf g},{\bf h},{\bf k})
\end{equation}
where
\begin{align}
     d\nu_3 &= e^{-2\pi i \mathcal{O}_4}, \\
     \mathcal{O}_4 &= \frac{1}{2} n_2(n_2 + \omega_2) + \frac{c_-}{8} \omega_2^2 \mod 1.
\end{align}
After plugging in the functional forms of $\omega_2, n_2$ and integrating, we get the following solution for $\nu_3$:
\begin{equation}
\begin{split}
    \nu_3({\bf g},{\bf h},{\bf k}) = &\exp\left(-2\pi i\left[\frac{k_s(k_{\pm}+k_s)}{2}+\frac{c_- k_{\pm}^2}{8}+k_3\right]\frac{[{\bf g}]_M}{M}\frac{[{\bf h}]_M+[{\bf k}]_M-[{\bf hk}]_M}{M}\right).
    \end{split}
\end{equation}
Here $k_3$ is a bosonic SPT index. Now a computation gives the desired result
\begin{equation}\label{eq:IM+-}
    \mathcal{I}^{\pm}_M = \exp\left(\frac{2\pi i}{M}\left[\frac{k_s(k_{\pm}+k_s)}{2}+\frac{c_- k_{\pm}^2}{8}+k_3\right]\right).
\end{equation}
Combining Eqs.~\eqref{eq:partialrotresult_CMM1},~\eqref{eq:partialrotresult_CMM(-1)F} with Eq.~\eqref{eq:IM+-}, and restoring the $\OO$ subscripts, we obtain Eq.~\eqref{eq:mainresultCFT} in the main text. Note that there is a phase proportional to $c_-$ coming from $\mathcal{I}^+_M$ but not $\mathcal{I}_M^-$. 

\section{Computation of $\ell_{s,\OO,\text{LL}}$}\label{app:LL}

Here we state the effective response theory for a continuum system of $C$ filled Landau levels with symmetry $\text{U}(1) \times \text{SO}(2)$ (where $\text{SO}(2)$ signifies spatial rotations), and find its relationship to the continuum limit of the Hofstadter model.

In terms of a $\text{U}(1)$ gauge field $A$ and an $\text{SO}(2)$ gauge field $\omega$ (which can be identified with the components of the spin connection on the underlying spatial manifold), the continuum response theory in the case where a $2\pi$ spatial rotation acts trivially on fermions is
\begin{equation}\label{eq:Leff_continuum}
    \mathcal{L}_{eff}=\sum_{n=1}^C( \frac{1}{4\pi} (A+s_n\omega) \wedge d(A+s_n\omega)-\frac{1}{48\pi}\omega \wedge d\omega)
\end{equation}
where $s_n=\frac{2n-1}{2}$ is the orbital spin per particle in the n-th Landau level. The Chern number of each filled Landau level is 1. The last term represents the gravitational anomaly. Expanding the above Lagrangian, we obtain

\begin{equation}
    \mathcal{L}_{eff}= \frac{1}{4\pi}A \wedge dA+ \frac{1}{2\pi}\frac{C^2}{2}A \wedge d\omega + \frac{1}{4\pi}\frac{2C^3-C}{6}\omega \wedge  d\omega.
\end{equation}
The coefficient of $\frac{1}{2\pi} A \wedge d\omega$ is identified with the Wen-Zee shift $\mathscr{S}^+_{\text{LL}}$. It has no origin dependence, and takes the value $\mathscr{S}^+_{\text{LL}} = C^2/2$. The coefficient $\tilde{\ell}^+_{s,\text{LL}}$ of the term $\frac{1}{4\pi}\omega \wedge d\omega$ also has no origin dependence, and takes the values 
\begin{equation}\label{eq:tildels_theory}
    \tilde{\ell}_{s,\text{LL}}^+=\frac{2C^3-C}{6}=\{\frac{1}{6},\frac{7}{3},\frac{17}{2},\frac{62}{3},\frac{245}{6},\dots\}
\end{equation}
We define $\ell_{\text{LL}} = \tilde{\ell}_s + C/12$, which is the coefficient of $\frac{1}{4\pi} \omega \wedge d \omega$ if we ignore the framing anomaly. 

In the continuum limit of the Hofstadter model, assuming that a $2\pi$ rotation acts trivially on fermions (that is, the rotation operator is given by $\Cmop$), the analogous response coefficient is denoted $\ell^+_{s,\OO,\text{LL}} = \ell^+_{s,\text{LL}} \mod \MO/2$; it acquires an origin dependence and is quantized mod $\MO/2$ where $\MO$ is the order of rotations that preserve $\OO$. Here we assume $\MO$ is even. For the square lattice,
\begin{equation}\label{eq:ells+oLL}
    \ell^+_{s,\OO,\text{LL}}=\frac{2C^3-C}{6}+\frac{C}{12} =\{\frac{1}{4},\frac{5}{2},\frac{3}{4},1,\frac{5}{4}\dots\}\mod \MO/2.
\end{equation}

To derive the expression for $\ell^-_{s,\OO,\text{LL}}$ when we instead use the operators $\Cmom$, we can simply replace $A \rightarrow A + \omega/2$ in Eq.~\eqref{eq:Leff_continuum} (this corresponds to choosing a reference rotation operator $\Cmom$ that inserts $\pi/\MO$ flux at a disclination of angle $2\pi/\MO$) and again compute the $\frac{1}{4\pi} \omega \wedge d\omega$ coefficient, after subtracting off the framing anomaly. The final result is
\begin{equation}\label{eq:ells-oLL}
    \ell^-_{s,\OO,\text{LL}}=\frac{2C^3-C}{6}+\frac{C^2}{2}+\frac{C}{4}+\frac{C}{12}=\{1,1,2,2,3,3,0,0,1,\dots\}\mod \MO.
\end{equation}
The above equations reproduce Eqs.~\eqref{eq:Theta+LL},~\eqref{eq:Theta-LL} appear in the main text.

\end{widetext}

\end{document}